\documentclass[aps,preprint,onecolumn,nofootinbib]{revtex4}

\usepackage{bm}
\usepackage{mathrsfs}
\usepackage{amssymb}
\usepackage{amsmath}
\usepackage{amsfonts}
\usepackage{color}
\usepackage{ulem}

\usepackage[]{graphicx}
\begin{document}
\newcommand{\half}{\frac{1}{2}}
\title{Collapse of the many-worlds interpretation: Why Everett's theory is typically wrong}
\author{Aur\'elien Drezet$^{1}$}
\address{$^1$Univ. Grenoble Alpes, CNRS, Grenoble INP, Institut Neel, F-38000 Grenoble, France}

\email{aurelien.drezet@neel.cnrs.fr}
\begin{abstract}
We analyze the objective meaning of probabilities in the context of the many-worlds interpretation of Everett. For this purpose we study in details the weak law of large numbers and the role of typicality and universally neglible probabilities (through the works of Cournot 	and Borel).  We demonstrate that Everett's theory doesnt provide any clue for fixing a  probability rule and therefore contradicts irrevocably empirical facts and Born's law.        
\end{abstract}

\maketitle
\section{Introduction and motivation}\label{sec1}
Everett's many-worlds interpretation \citep{Everett1957}, abbreviated as MWI in the following, is certainly one of the most known and discussed attempt for finding an alternative ontology of quantum mechanics. Unlike the standard Copenhagen or `collapse' interpretation the MWI is based on a purely unitary ontology based on the Schr\"odinger equation and doesn't require the unfamous, and vaguely defined, separation between observers and observed systems.\\
\indent The motivations for developing the MWI were clearly announced by Everett~\citep{Everett1957,Barrett2012,DeWitt1973}. On the one side, we want a complete theory describing the Universe as a whole and without an arbitrary `shifty split' \`a la Heisenberg-von Neumann between different levels of reality. On the other side, we also want the theory to be able of recovering the many statistical features and successes of the orthodox/Copenhagen interpretation. Most importantly, we need to recover the usual Max Born probability law
\begin{eqnarray}\mathcal{P}^{(\textrm{Born})}_\alpha=|\Psi_\alpha|^2=|\langle\alpha|\Psi\rangle|^2
\end{eqnarray} for observing an outcome $\alpha$ after a quantum measurement acting upon the state $|\Psi\rangle=\sum_\alpha\Psi_\alpha |\alpha\rangle$.\\ 
\indent Recovering Born's rule and the statistical interpretation has been claimed by DeWitt to constitute the EWG (Everett-Wheeler-Graham) `meta-theorem': 
\begin{quote}
\textit{The mathematical formalism of the quantum theory is capable of yielding its own interpretation.}~\citep{DeWitt1971,DeWitt1973} 
\end{quote} 
This `proof' (and its descendants~\citep{Hartle1968,Ochs1979,Geroch1984,Farhi1989,Ohkuwa1993,Coleman1994,Gutmann1995,Tegmark1998,Aharonov2002,Aguire2011,Wesep2006}) has been strongly criticized over the years and accused of circularity in particular because it relies on many subtleties concerning the application of the Bernoulli law of large-numbers, e.g.,  concerning very long sequences of repeated experiments~\citep{Ballentine1973,Benioff1978,Stein1984,Kent1990,Squires1990,Putnam2005,Hemmo2007,Adlam2014,Barrett2017}.\\
\indent  More recently, the mere idea that a deterministic ontology based on Schr\"odinger's unitary equation could lead to something like a justification of the probabilistic properties of our Universe has been criticized and named the `incoherence problem' by philosophers~\citep{bookMWI,Albert2015,Maudlin2019}. Indeed, in the MWI all the alternatives of a quantum measurement acting upon $|\Psi\rangle=\sum_\alpha\Psi_\alpha |\alpha\rangle$ are actual: None of them are collapsing and all of them are required to survive in order to preserve the unitary time-evolution of the whole universal wave-function. In this context, how could we with the MWI even hope to justify stochasticity, i.e., as observed in the lab? Nevertheless, several authors have attempted to answer repeated objections made against the MWI and have tried to `precise' the definition of probabilities used in this theory. For instance, it has often been claimed, in reply to criticisms, that probabilistic concepts used in the MWI are not worse (and perhaps not better) than they are in other interpretations of quantum mechanics or even in classical statistical mechanics \citep{Everett1957,DeWitt1971,Papineau1996,Wallace2012}. In parallel, Vaidman~\citep{Vaidman1998,Vaidman2012,Vaidman2014,Vaidman2020,McQueen2019}, partly in response to Albert \citep{bookMWI,Albert2015}, has developped suggestive narratives and a complete semantics for speaking about probability and `self-locating uncertainty' as understood and perceived by observers in quantum branches of the universal wavefunction (see also~\citep{Greaves2004,Tappenden2010,Tappenden2019}). One of the proposed solution advocated by McQueen and Vaidman~\citep{Vaidman1998,Vaidman2012,Vaidman2014,Vaidman2020,McQueen2019} is to assume an additional postulate for the MWI: the so-called Born-Vaidman rule~\citep{Tappenden2010} 
\begin{quote}
\textit{MWI probability postulate:  The probability of self-location in a world with a given set of outcomes is the absolute square of that world's amplitude.}~\citep{McQueen2019} 
\end{quote}  As stressed by these authors `the postulate is not about some fundamental physical process, it is about the experience of the observer'. This is a subjective definition of probability that would probably not reject de Finetti himself~\citep{Finetti1937}. Moreover, accepting a objective-Bayesianist reading of probability and Lewis's `principal-principle', it must also be assumed that the degree of belief or credence $\mathcal{C}_\alpha$, for the observer to be in one branch, is tied to an objective chance $\mathcal{P}_\alpha$ \footnote{More precisely we write $\mathcal{C}_\alpha:=\mathcal{C}(\alpha|\mathcal{P}_\alpha)=\mathcal{P}_\alpha$~\citep{Lewis1980}.}. Following Vaidman, this objective property is just the `measure of existence' $\mathcal{P}_\alpha=|\Psi_\alpha|^2$, i.e., as given by Born's rule. This supplementary rule is needed in order to constraint the statistics in the MWI to follow empirical evidences like Born's law. Yet, this strategy constitutes an amendment to the original Everett goal and not every advocates of the MWI subscribe to it.  \\ 
\indent Furthermore, several new ideas based on decision-theoretic scenario  \`a  la Deutsch~\citep{Deutsch1999,Wallace2007,Saunders2005,Saunders2008,Wallace2012} or `envariance' \`a la Zurek \citep{Zurek2003,Zurek2005,Barnum2003,Sebens2016} have been discussed as alternative `proofs' for making sense of probability in the MWI, i.e., purportedly without incoherence and with the correct Born's rule. These ideas have also been strongly criticized~\citep{bookMWI,Albert2015,Maudlin2019,Kent2015} because they rely on some purely personalist and Bayesian definitions of probability that are, \emph{apriori}, unrelated to experimental facts (i.e, relative frequencies). In this decision-theoretic `Oxfordian' approach it is only proved that \emph{if} we could define a probability function in quantum mechanics  this function will satisfy Born's rule. However, the physical explanation of the probability assumption (i.e., the qualitative or incoherence problem) is standing unjustified and apparently leads to circular deductions if we dont add something to the MWI~\citep{Adlam2014,Drezet2016}. \\
\indent In the present work, we are not going to further discuss the meaning of subjective probabilites in the decision-theoretic and Bayesian scenarios.  Instead, we are going back to the old EWG theorem which is still we think at the core of the problem. More precisely, we believe that it is only by going back to the classical definitions of a probability as given in the work of Bernoulli, Laplace, Borel and many others that one can understand the objective and pragmatic  role of probabilities as estimators of relative frequencies (even though if all these authors also emphasized the subjective role of probabilities). We will show that the mathematical and physical definitions of probabilities, and particularly the weak law of large numbers (WLLN), constitute actually fatal issues for the MWI. It is therefore, we think, very important to understand the consequences and the unavoidability of the problem. In turn, it will we think demonstrate that the MWI can just not be true in its basic form. It implies that  one must either modify the ontology (like for instance in the many de Broglie-Bohm worlds strategy~\citep{Tipler2014,Sebens2014,Bostrom2014}), or add something to the current MWI as suggested  by Vaidman (the present author has recently speculated a toy model of that kind \citep{Drezet2021} based on a unitary version of the many-minds interpretation~\citep{AlbertLoewer1988}).\\ 
\indent Of course, these are very strong claims and here we propose to justify this view with a very elementary scenario using a Bernouilli sequence of a simple quantum coin tossing (an elementary description adapted to the MWI is given in section \ref{sec2}). During the discussion we will in section \ref{sec3} study the history of the WLLN that is central to undertand the MWI. Furthemore, we will discuss the notions of typicality and universally negligible probabilities that are needed to decipher the meaning of that WLLN.  In particular, we will in section \ref{sec4} bring new elements coming from the deterministic de Broglie-Bohm theory or pilot-wave interpretation (PWI)~\citep{BohmHiley}, that is an alternative hidden-variables quantum theory, in order to clarify the concept of probability involved in the MWI and show that the full Everettian theory can just not survive in its present form.              
\section{Quantum Bernoulli process}\label{sec2}
\indent We start with a two-level quantum system or qubit described by the wave-function
\begin{eqnarray}
|\Psi_1\rangle=a|\spadesuit\rangle+b|\heartsuit\rangle\label{eq1} 
\end{eqnarray} with $a,b\in \mathbb{C}$ and $\langle\heartsuit|\spadesuit\rangle=0$, $\lVert|\spadesuit\rangle\rVert=\lVert|\heartsuit\rangle\rVert=1$ (we also impose the normalization $|a|^2+|b|^2=1$).\\ 
\indent Now, we suppose a quantum experiment involving a kind of Stern-Gerlach apparatus to separate the two components $\spadesuit/\heartsuit$  of the system. After this splitting (that involves a form of external field acting on the qubit) we have not yet a quantum measurement.  In the standard `collapse' interpretatation we require the separated wave-packets associated with the two components $\spadesuit/\heartsuit$  to interact with amplifying devices or detectors bringing the information from the microscopic quantum world of potentiality to the macroscopic world where only one outcome has been registered. \\ 
\indent However, in the MWI we preserve unitarity and we must include everything, i.e., even the observers, into the quantum realm. Indeed, the awesome idea of Everett~\citep{Everett1957,Barrett2012,DeWitt1973} was to introduce observers as quantum mechanical devices participating to the experiment. As Everett wrote in his doctoral thesis:       
\begin{quote}
\textit{As model for observers we can, if we wish, consider automatically functioning  machines, possesing sensory apparata and coupled to recording devices capable of registering past sensory data and machine configurations.}~\citep[p.~64]{DeWitt1973}
\end{quote} 
Here, before the measurement (lets say at time $t_0$) we write  $|\mathcal{M}_0,\mathcal{A}\textrm{ur\'elien}_0\rangle_{t_0}$ the joint quantum state involving the detector $\mathcal{M}$, the observer and his environement~\footnote{These sub-systems are strongly entangled, e.g., due to decoherence, and therefore our notations don't even try to clearly identify and separate the Hilbert space belonging to the observer from the one associated with the apparatus or with the environement.}. The interaction or measurement process that we consider has several important steps summarized on the quantum state evolution
\begin{eqnarray}
|\Psi_1\rangle_{t_0}\otimes|\mathcal{M}_0,\mathcal{A}\textrm{ur\'elien}_0\rangle_{t_0}\nonumber\\
\rightarrow (a|\tilde{\spadesuit}\rangle_{t_1}+b|\tilde{\heartsuit}\rangle_{t_1}) \otimes|\mathcal{M}_0,\mathcal{A}\textrm{ur\'elien}_0\rangle_{t_1}\label{eq2_1}\\ 
\rightarrow a|\mathcal{M}^\ast_\spadesuit,\mathcal{A}\textrm{ur\'elien}_\spadesuit\rangle_{t_2} +b|\mathcal{M}^\ast_\heartsuit,\mathcal{A}\textrm{ur\'elien}_\heartsuit\rangle_{t_2}\label{eq2_2}\\ 
\rightarrow a|\mathcal{M}_0,\mathcal{A}\textrm{ur\'elien}_\spadesuit\rangle_{t_3} +b|\mathcal{M}_0,\mathcal{A}\textrm{ur\'elien}_\heartsuit\rangle_{t_3},\label{eq2_3}
\end{eqnarray} where we have added a time label to follow the chronology. At time $t_1$, i.e., Eq.~\ref{eq2_1}, the splitting of the initial wave packet $|\Psi_1\rangle_{t_0}$ already occurred: we obtain two disjoints and orthogonal wave packets $|\tilde{\spadesuit}\rangle_{t_1}$ and $|\tilde{\heartsuit}\rangle_{t_1}$ (i.e., $\langle\tilde{\heartsuit}|\tilde{\spadesuit}\rangle_{t_1}=0$, $\lVert|\tilde{\spadesuit}\rangle_{t_1}\rVert=\lVert|\tilde{\heartsuit}\rangle_{t_1}\rVert=1$) moving in the direction of the detectors. At that time the detectors and observer are still factorized from the qubit state~\footnote{Note that $|\mathcal{M}_0,\mathcal{A}\textrm{ur\'elien}_0\rangle_{t_1}$ is generally different from $|\mathcal{M}_0,\mathcal{A}\textrm{ur\'elien}_0\rangle_{t_0}$ but that unitarity imposes that the norm is preserved: $\lVert|\mathcal{M}_0,\mathcal{A}\textrm{ur\'elien}_0\rangle_{t_1}\rVert=\lVert|\mathcal{M}_0,\mathcal{A}\textrm{ur\'elien}_0\rangle_{t_0}\rVert=1$.}.\\ 
\indent At time $t_2$, i.e., Eq.~\ref{eq2_2}, the detectors have detected and absorbed the qubit (the notation $\mathcal{M}^\ast$ means that the devices are excited). Rigorously, this involves two detectors: one being located along the trajectory of the $\tilde{\spadesuit}$ wave packet and one along the path of the  $\tilde{\heartsuit}$ wave packet (i.e., we have here a kind of coincidence measurement like the one involved in Hanbury Brown and Twiss interferometers for single photons). We have  $\langle\mathcal{M}^\ast_\heartsuit,\mathcal{A}\textrm{ur\'elien}_\heartsuit|\mathcal{M}^\ast_\spadesuit,\mathcal{A}\textrm{ur\'elien}_\spadesuit\rangle_{t_2}=0$ and $\lVert|\mathcal{M}^\ast_\spadesuit,\mathcal{A}\textrm{ur\'elien}_\spadesuit\rangle_{t_2}\rVert=1$, $\lVert|\mathcal{M}^\ast_\heartsuit,\mathcal{A}\textrm{ur\'elien}_\heartsuit\rangle_{t_2}\rVert=1$ as required by unitarity.\\  
\indent At the final stage $t_3$, i.e., Eq.~\ref{eq2_3}, the two detectors have relaxed to their ground states and the environement has kept an information about the process which for simplicity is included in the state of the observer. Once again, unitarity imposes that we get:  $\langle\mathcal{M}_0,\mathcal{A}\textrm{ur\'elien}_\heartsuit|\mathcal{M}_0,\mathcal{A}\textrm{ur\'elien}_\spadesuit\rangle_{t_2}=0$ and $\lVert|\mathcal{M}_0,\mathcal{A}\textrm{ur\'elien}_\spadesuit\rangle_{t_2}\rVert=1$, $\lVert|\mathcal{M}_0,\mathcal{A}\textrm{ur\'elien}_\heartsuit\rangle_{t_2}\rVert=1$. We finally obtain two orthogonal universes in which the two copies of the observer unaware of each other are living with their own memories of the experimental result.\\  
\indent We stress that the coefficients $|a|^2$ and $|b|^2$, that are in the orthodox interpretation identified with the Born probabilities for observing  $\spadesuit$ or $\heartsuit$, i.e., \begin{eqnarray}
\mathcal{P}_\spadesuit=|a|^2 \textrm{ or } \mathcal{P}_\heartsuit=|b|^2, \label{heart}
\end{eqnarray} are here not yet identified with probabilities, but  merely represent intensities or if you wish `measures of existence' \citep{Vaidman2012,Vaidman2014} for the two orthogonal branches.\\ 
\begin{figure}[h]
\begin{center}
\includegraphics[width=8.5cm]{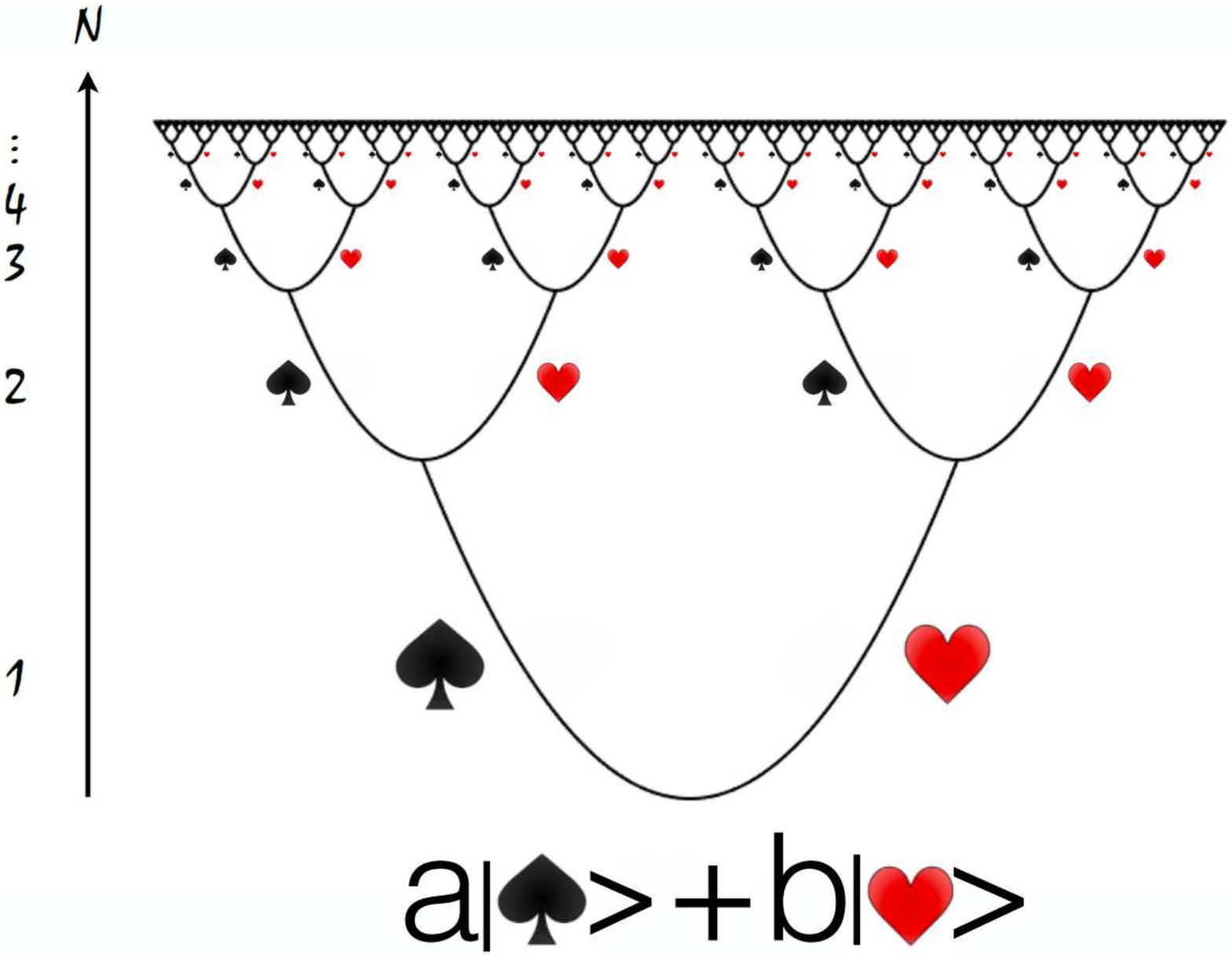} 
\caption{Schematic representation as a fractal tree of the quantum Bernoulli process associated with the quantum state $|\Psi_1\rangle=a|\spadesuit\rangle+b|\heartsuit\rangle$. } \label{figure1}
\end{center}
\end{figure}
\indent The next step in our analysis is to consider a long sequence of  $N$ trials of the same experiment (i.e., a Bernoulli process as seketched in Figure \ref{figure1}). We begin with a product state that reads $|\Psi_N\rangle=\bigotimes_{i=1}^{N}|\Psi_1^{(i)}\rangle:=|\Psi_1^{(1)}\rangle\otimes...\otimes|\Psi_1^{(N)}\rangle$ where, as in Eq.~\ref{eq1}, we have $|\Psi_1^{(i)}\rangle=a|\spadesuit^{(i)}\rangle+b|\heartsuit^{(i)}\rangle$ ($i$ labeling the different Hilbert spaces associated with the different qubits). Each of the $N$ qubits are successively analyzed using our Stern-Gerlach apparatus and the sequence of results is carefully registered and memorized by the observer.  In the end, we obtain a sum of $2^N$ orthogonal decohered branches or universes which reads 
 \begin{eqnarray}
|\Psi_N\rangle_{t_0}\otimes|\mathcal{M}_0,\mathcal{A}\textrm{ur\'elien}_0\rangle_{t_0}\nonumber\\ 
\rightarrow \sum_h a^{N^h_\spadesuit} b^{N^h_\heartsuit}|\mathcal{M}_0,\mathcal{A}\textrm{ur\'elien}_h\rangle_{t_f}.\label{eq3}
\end{eqnarray}
 In Eq.~\ref{eq3}, $h:=[\alpha^h_1,\alpha^h_2,...,\alpha^h_N]$ with $\alpha^h_i= \spadesuit$ (or $\heartsuit$) denotes one history in which the $i^{\textrm{th.}}$ measurement led to the outcome $|\spadesuit^{(i)}\rangle$ (or $|\heartsuit^{(i)}\rangle$).\\ 
 \indent To each history $h\in \mathcal{H}$ (where $\mathcal{H}:=\{\spadesuit,\heartsuit\}^N$ is the ensemble of all histories with cardinality $2^N$) is associated a quantum observer state $|\mathcal{M}_0,\mathcal{A}\textrm{ur\'elien}_h\rangle_{t_f}$. These states constitute a complete set such that for each possible pair of histories $h,h'$ we have $\langle\mathcal{M}_0,\mathcal{A}\textrm{ur\'elien}_{h'}|\mathcal{M}_0,\mathcal{A}\textrm{ur\'elien}_h\rangle_{t_f}=\delta_{h,h'}$ (where $\delta_{h,h'}$ is a Kronecker symbol). \\ 
\indent Moreover, the important quantity in Eq.~\ref{eq3} is the branch amplitude $a^{N^h_\spadesuit} b^{N^h_\heartsuit}$ which involves the number of times $N^h_\spadesuit$ (respectively $N^h_\heartsuit$) that a $\spadesuit$ (respectively $\heartsuit$) appeared in the sequence $h$ of length $N$.  We have $N^h_\spadesuit+N^h_\heartsuit=N$, and by definition of $h$ we also have $N^h_\alpha=\sum_{i=1}^{i=N}\delta_{\alpha,\alpha^h_i}$ where $\alpha= \spadesuit$ or $\heartsuit$. It is convenient to introduce the relative frequency of occurence~\citep{DeWitt1973,Hartle1968} for the outcome $\alpha$ in the history $h$ as 
 \begin{eqnarray}
  Q^h_\alpha:=\frac{N^h_\alpha}{N}=\frac{1}{N}\sum_{i=1}^{i=N}\delta_{\alpha,\alpha^h_i}
  \end{eqnarray} 
with clearly  $Q^h_\spadesuit+Q^h_\heartsuit=1$. All these concepts are required for the statistical analysis discussed in the next section.
\section{The law of large numbers: probability and typicality}\label{sec3} 
\indent In order to discuss the concept of probablities we  consider first the previous Bernoulli process from the point of view of the Copenhagen interpretation (or if we wish any collapse or reduction model of the quantum world~\citep{GRW}).\\  
\indent More precisely, from the point of view of a super-observer localized on the classical side of the quantum-classical boundary and watching the full experiment described by Eq.~\ref{eq3} we know that $\mathcal{A}_h=a^{N^h_\spadesuit} b^{N^h_\heartsuit}$ represents a `probability amplitude' for the alternative $h$. Therefore, applying Born's rule and Eq.~\ref{heart} we obtain the probability $\mathcal{P}_h$ for the sequence $h$ as 
\begin{eqnarray}
      \mathcal{P}_h=|\mathcal{A}_h|^2=\mathcal{P}_\spadesuit^{N^h_\spadesuit}\mathcal{P}_\heartsuit^{N^h_\heartsuit}. \label{probh}
\end{eqnarray}The concept of probability that is used here is still vague and its relation to the experimental world will be precised in the following in connection with the works of Laplace and others.\\           
\indent First, observe that for such a sequence of Bernoulli trials we are not interested in $\mathcal{P}_h$ but instead in the sum \begin{equation}\sum_{h\in \mathcal{H}_{N_\spadesuit}} \mathcal{P}_h:=\mathcal{P}(N_\spadesuit)\end{equation} in the subset $\mathcal{H}_{N_\spadesuit}$ of histories corresponding to a fixed value of $N_\spadesuit$ (and therefore $N_\heartsuit=N-N_\spadesuit$). From pure combinatorics we immediately deduce the binomial probability for recording a number $N_\spadesuit$ of $\spadesuit$ in a sequence of $N$ independent experiments:
\begin{eqnarray}
      \mathcal{P}(N_\spadesuit)=\frac{N!}{N_\spadesuit!N_\heartsuit!}\mathcal{P}_\spadesuit^{N_\spadesuit}\mathcal{P}_\heartsuit^{N_\heartsuit}
      \simeq \frac{e^{-\frac{1}{2}\left(\frac{Q_\spadesuit-\mathcal{P}_\spadesuit}{\delta Q_\spadesuit}\right)^2} }{N\delta Q_\spadesuit\sqrt{2\pi}}\label{laplace}
\end{eqnarray}  where the second equality results from the Moivre-Laplace central-limit theorem valid if $N\gg 1$ (i.e., after a direct application of Stirling formula). Here, the probability $\mathcal{P}_\spadesuit$ is identified with the mean value $\langle Q_\spadesuit\rangle$ of the random variable $Q_\spadesuit=N_\spadesuit/N$ and similarly $\sqrt{\left(\frac{\mathcal{P}_\spadesuit\mathcal{P}_\heartsuit}{N}\right)}$ with the fluctuation  $\delta Q_\spadesuit=\sqrt{\langle (Q_\spadesuit-\langle Q_\spadesuit\rangle)^2\rangle}$.\\ \indent Most importantly for the present discussion, mathematics shows that for any real number $\sigma\in[0,\frac{\textrm{min}(\mathcal{P}_\spadesuit,\mathcal{P}_\heartsuit)}{\delta Q_\spadesuit}]$  the probability $\mathcal{P}(|Q_\spadesuit-\mathcal{P}_\spadesuit|\geq \sigma\delta Q_\spadesuit)=\sum_{|Q_\spadesuit-\mathcal{P}_\spadesuit|\geq \sigma\delta Q_\spadesuit} \mathcal{P}(N_\spadesuit)$ admits the upper bound:
\begin{eqnarray}
   \mathcal{P}(|Q_\spadesuit-\mathcal{P}_\spadesuit|\geq \sigma\delta Q_\spadesuit)\leq 2e^{-2N\sigma^2\delta Q^2_\spadesuit}=2e^{-2\sigma^2\mathcal{P}_\spadesuit\mathcal{P}_\heartsuit}   \label{err} 
\end{eqnarray} known as Hoeffding's bound~\footnote{\label{foot1}More generally, in the continuous limit where $N\gg 1$ we have: \begin{eqnarray}
   \mathcal{P}(|Q_\spadesuit-\mathcal{P}_\spadesuit|\leq \sigma\delta Q_\spadesuit)\simeq \frac{1}{\sqrt{2\pi}}\int_{-\sigma}^\sigma dxe^{-\frac{x^2}{2}}:=\textrm{erf}(\frac{\sigma}{\sqrt{2}})   \label{err2}
\end{eqnarray} which leads to the famous 3-$\sigma$ rule corresponding to $\sigma=3$ in Eq.~\ref{err}, i.e., $\mathcal{P}(|Q_\spadesuit-\mathcal{P}_\spadesuit|\leq 3\delta Q_\spadesuit)\simeq 99.73\% $. For $\sigma\gg 1$ we can use the asymptotic formula $\textrm{erf}(\frac{\sigma}{\sqrt{2}})\simeq 1-\frac{e^{-\sigma^2/2}}{\sigma\sqrt{2\pi}}$.}. After writing  $\varepsilon =\sigma\delta Q_\spadesuit$ Eq.~\ref{err2} is used to obtain a version of the (weak) law of large numbers (WLLN):
\begin{eqnarray}
  \lim_{N \to +\infty} \mathcal{P}(|Q_\spadesuit-\mathcal{P}_\spadesuit|\geq \varepsilon)=0, & \forall \varepsilon>0 \label{err3}
\end{eqnarray} showing that in the $N \rightarrow +\infty$ limit $Q_\spadesuit$ converges probabilistically to $\mathcal{P}_\spadesuit$~\footnote{Note that the WLLN is more often deduced from the Bienaym\'e–Chebyshev inequality $\mathcal{P}(|Q_\spadesuit-\mathcal{P}_\spadesuit|\geq \varepsilon)\leq \delta Q_\spadesuit^2/\varepsilon^2$. Note also, that in the continuous limit Eq.~\ref{laplace}  implies  that   $\mathcal{P}(aN\leq N_\spadesuit\leq bN)\simeq\int_a^b dQ\rho_{\delta Q_\spadesuit}(Q-\mathcal{P}_\spadesuit)$ where the Gaussian probability distribution \begin{eqnarray}
\rho_{\delta Q_\spadesuit}(X)=\frac{1}{\delta Q_\spadesuit\sqrt{2\pi}}e^{-\frac{1}{2}\left(\frac{X}{\delta Q_\spadesuit}\right)^2}
\end{eqnarray} asymptotically approaches the Dirac distribution $\delta(X)$ in the $N \rightarrow +\infty$ limit.}. This results which was originally obtained by Jacob Bernoulli around 1692  is often considered as a `golden-theorem' of the probability calculus.\\ 
\indent Moreover, the WLLN is used to show that for `almost all' alternatives $h\in\mathcal{H}$ the frequency $Q_\spadesuit$ (respectively $Q_\heartsuit$) is a very good estimator of the probability $\mathcal{P}_\spadesuit$ (respectively $\mathcal{P}_\heartsuit$) when $N$ is very large. By almost we precisely mean that histories for which Born's rule $Q_\spadesuit\simeq|a|^2$, $Q_\heartsuit\simeq|b|^2$  holds true are `typical' in probability.\\
\indent At that level of our discussion we must now try to introduce a definition of probablity. We remind that the classical definition of a probability~\citep{Hacking1975,Stigler1986} given by Laplace is 
\begin{quote}
\textit{ One has seen in the introduction that the probability of an event is the ratio of the number of cases that are favourable to it, to the number of possible cases, when there is nothing to make us believe that one case should occur rather than any other, so that these cases are, for us, equally possible. }~\citep[p.~181]{Laplace}.
\end{quote} A similar definition was already given by de Moivre in 1718:
 \begin{quote}
\textit{ The probability of an event is greater or less, acording to the number of chances by which it may happen, compared with the number of all the chances, by which it may either happen or fail.  }~\citep[p.~1]{deMoivre1718}.
\end{quote}
There are several parts in the de Moivre-Laplace definition but let us here focus on this counting or enumeration procedure for `chances' i.e., physically distinct   alternatives. Moreover, contrarily to Laplace we dont accept an ignorance-based justification for probabilities in a deterministic Universe.  To paraphrase Poincar\'e: information given by the probability calculus will not stop to be true  the day when these phenomena will be better known \citep[p.~3]{Poincare1912}. Therefore, we assume that the so-called `subjective probabilities' are only estimations of (objective) probabilities. The central question is of course which objective meaning can have the word  chance in a purely deterministic Universe if we dont want to introduce some new mysterious and occult fluid (see the remark in the footnote \ref{remarkus}). We should try to answer this question in the following.   Moreover, accepting the `classical' Moivre-Laplace definition the probability $\mathcal{P}(A)$  for an outcome or event $A$ is given  by the ratio \begin{eqnarray}
\mathcal{P}(A):=\frac{M(A)}{M}\label{Laplacedef}
\end{eqnarray} 
between the number of elementary alternatives $M(A)$ leading to the outcome $A$ and the total number of alternative  $M$ in the game.\\
\indent Now, of course counting alternatives or chances is not without ambiguities.  Take a coin (classical or quantum): we have only two elementary alternatives so that apriori probabilities for elementary outcomes $\spadesuit$ and $\heartsuit$ are necessarily $\mathcal{P}_\spadesuit=\mathcal{P}_\heartsuit=\frac{1}{2}$. Moreover, in order to extend or apply Laplace's definition to the cases of  unfair coins one must go beyond simple `geo-metrical' considerations. Consider again our coin with two alternatives $\spadesuit,\heartsuit$, and now suppose that each alternative is actually degenerated and divided in many `physically available' sub-alternatives $M_\spadesuit$ and $M_\heartsuit$ respectivelly. The classical definition now applies and allows for any non-negative (rational number) probabilities $\mathcal{P}_{\spadesuit/\heartsuit}= \frac{M_{\spadesuit/\heartsuit}}{M_{\spadesuit}+M_{\heartsuit}}$ with the normalization $\mathcal{P}_\spadesuit+\mathcal{P}_\heartsuit=1$. Now, it seems that we actually introduced mysterious new features in the formalism. What are actually those internal degrees of freedom  with numbers 
$M_\spadesuit$ and $M_\heartsuit$? By introducing such elements we actually defined a new metric or weight on the configuration space $\{\spadesuit,\heartsuit\}$ with cardinality 2.  An alternative way to see that is to consider that each elementary physical alternative $\spadesuit$, or $\heartsuit$ corresponds to a urn filled with $M_\spadesuit$ and $M_\heartsuit$ distinguishable balls or particles. When $M\rightarrow +\infty$ we have with this atomic model a way to picture and approximate any continuous non negative distributions of probability $\mathcal{P}_{\spadesuit/\heartsuit}\in [0,1]$ with $\mathcal{P}_\spadesuit+\mathcal{P}_\heartsuit=1$. In others words, we have here progressively building the representation of a continuous fluid that is reminiscent of  the definition made by Gibbs of a `fictitious' probability fluid or ensemble~\footnote{Of course, the Gibbs ensembles were historically not built on a discrete space $\{\spadesuit,\heartsuit\}$ but on a continuous phase-space $\Gamma\in \mathbb{R}^{6r}$ where $r$ is an integer and $6r$ the number of degrees of freedom of the ensemble (not to be confused with the cardinality  $\textrm{card}\left(\Gamma\right)=(2^{\aleph_0})^{6r}=2^{\aleph_0}$).}. More precisely, Gibbs following Maxwell and Boltzmann wrote \begin{quote}
\textit{ We may imagine a great number of systems of the same nature, but differing in the configurations and velocities which they have at a given instant, and differing not merely infinitesimally, but it may be so as to embrace every conceivable combination of configuration and velocities.}\citep[p.~vii]{Gibbs1902}
\end{quote} This conception of a `great number of systems' is clearly reminiscent of Maxwell/Boltzmann gas theory in which huge numbers $M\simeq 10^{23}$ of molecules or atoms are considered.  Yet, for Gibbs this frequentist analogy is not the end ot the story since the word `imagine' emphasizes the mental~\footnote{Schr\"odinger~\citep[p.~3]{Schrodinger1952} uses the expression `mental copies' to denote a Gibbs ensemble.} or at least not actual nature of this ensemble. For Gibbs this is just a physical model for building a quasi-continuous probability fluid. In the end, the Gibbs atomic model (like the mechanical analogies proposed by Maxwell for modeling the Aether) is not required anymore and can be eliminated: we have obtained the general notion of a non-negative `measure',`weight' or `metric' $\mu(A)$ for a set of events $A$, i.e.,  as it was later formally axiomatized by Borel, Lebesgue and Kolmogorov at the beginning of 20$^{th}$ century.\\
\indent Following Laplace and thus Gibbs, the `confidence level' probability $\mathcal{P}(|Q_\spadesuit-\mathcal{P}_\spadesuit|\leq \sigma\delta Q_\spadesuit)$ must be interpreted as a fraction $\frac{M_{\mathcal{H}_{\textrm{typical}}}}{2^N}$ of the number of typical histories over the total number of alternatives, i.e., here $2^N$. The WLLN states that when $N$ is growing this ratio $\frac{M_{\mathcal{H}_{\textrm{typical}}}}{2^N}$ also grows and tends to the maximum value $\frac{2^N}{2^N}=1$ asymptotically. With this property, and assuming a precisely defined probability measure on the whole history set $\mathcal{H}$, overwhelmingly all possible histories $h\in\mathcal{H}$ ultimately belong to the typical history set $\mathcal{H}_{\textrm{typical}}\sim \mathcal{H}$.\\ 
 \indent For all practical purposes this means that by watching a long  Bernoulli sequence   with $N\gg 1$ (for example $N\sim 10^{10}$) the frequency $Q_\spadesuit$ deviates negligeably from the characteristic frequency, i.e., the probability $\mathcal{P}_\spadesuit$,  for almost all eventualities and possibilities $h$. Of course, this doesn't mean that an atypical event is forbidden:  There is nothing in the dynamics that prohibits an atypical event to happen (this would contradict the mere definition of a Bernoulli process).\\
\indent Importantly, it is by applying this WLLN that we make sense of the usual frequentist definition of a probability: 
\begin{eqnarray}
  \lim_{N \to +\infty}   Q^h_\alpha\equiv\mathcal{P}_\alpha \label{fre}
\end{eqnarray} where $h$ belongs to the typical set $\mathcal{H}_{\textrm{typical}}$ tending probabilistically to the whole set $\mathcal{H}$ in the $N\rightarrow +\infty$ limit. Once again, we stress~\citep{Borel1909,Borel1914,Cantelli1935,Kolmogorov1950}
 that this concept of `limit in probability'  based on typicality is different from the more usual point-wise limit $\lim_{N \to +\infty}   Q^h_\alpha=\mathcal{P}_\alpha $. Here, the notion of limit in probability is conditionned on the existence of a typical set $\mathcal{H}_{\textrm{typical}}$ itself defined probabilistically. This key point, often neglected, implies that the WLLN can not be used to define the probability as a frequency without further hypothesis or physical considerations on the notion of limit and sequences.\\
\indent  This has led to long standing debates concerning the physical relation between frequencies (i.e., statistics) and probabilities (i.e., possibilities). In particular, whereas the mathematical condition  Eq.~\ref{fre} recovering probabilities as relative frequencies in the limit $N\rightarrow +\infty$ is apriori not questionned, it appears nevertheless unphysical since related to an infinitely long Bernoulli sequence which is aposteriori  never encountered in nature.  Moreover, for finite $N\gg 1$ the typicality reasoning leading to the WLLN is also often criticized for being circular, i.e. leading to an infinite regress fallacy. To quote David Wallace:
\begin{quote}
\textit{ We cannot prove  that in the long-run relative frequencies converge to probabilities. What we can prove, is that in the long-run relative frequencies converge to probabilities...  Probably.}~\citep{Wallacevideo}
\end{quote} Furthermore, in the short-run the relation between relative frequencies and probabilities disappears and as a direct consequence the physical meaning of probabilities seems to vanish aswell. In other words, probabilities can not simply, i.e., with certainty, be identified with long-run frequencies without clearly understanding what is the physical meaning of typicality. \\ 
\indent We stress that the intuitive idea to define probability as frequency, i.e., `frequentism' has been defended to some extent by Ellis, Cournot and Venn~\citep{Ellis1843,Cournot1843,Venn1876} in the  19$^{th}$ century, and was later supported by von Mises~\citep{vonMises1957} and Reichenbach~\citep{Reichenbach1949} who considered infinite collectives. This interpretation of probability is often advocated by statisticians and physicists~\citep{Stigler1986} to justify the existence of statistical ensembles (e.g., in statistical mechanics~\citep{Reif1965}). Moreover, frequentism has also been strongly criticized by those who, like  Keynes~\citep{Keynes1921}, Ramsey~\citep{Ramsey1990} and de Finetti~\citep{Finetti1937}, prone a more subjectivist and personalist perspective on probability~\footnote{\label{remarkus} Many authors believe that determinism necessarily implies subjective probabilities.  Popper analyzed this psychological prejudice explaining that `\textit{it is in a way the only reasonable possibility which they can accept: for objective physical probabilities are incompatible with determinism; and if classical physics is deterministic, it must be incompatible with an objective interpretation of classical statistical mechanics}'~\citep{Popper1982}. Popper however  deliberately neglected a solution that he considered as conspiratory, i.e., assuming that initial conditions are prepared in the correct way in order to reproduce statistics and the WLLN. This `conspiracy' is actually the way we follow in the PWI in order to  define objective probabilities  without relying on ignorance \`a la Laplace  and without introducing `propensities' \`a la Popper and therefore without abandonning determinism. Moreover, the only empirical manifestation of propensities rely on statistics. Therefore, we assume  that such a disparate `magical' remedy advocated by Popper looks similar to introducing a new `phlogistic' occult substance for giving some materiality to `chances'. }. Sometimes, mathematicians like Borel~\citep{Borel1939} accept prudently a more nuanced and `rationalist' approach mixing subjectivism and frequentism (this view goes back at least to Laplace and is also named objective-Bayesianism~\citep{Jeffreys1961,Jaynes1999,Bricmont2016}). Repeated objections done against frequentism include the `single-case' and `reference class' problems (for a recent and interesting review of these objections against pure frequentism see \citep{Hajek1996,Hajek2009}) that are both connected to the interpretation of a probability assigned to a singular event. Here, we dont subscribe to the frequentism advocated by von Mises and others. Better, as we will show, we assume an objective interpretation of probability based on the concept of typicality. In this view frequencies are still fundamental but unlike von Mises we dont require infinite collectives and an alternative axiomatics to define probability.\\       
\indent To further understand and decipher the central role played by typicality in an objective approach to probability it is important to realize how efficient are these probabilistic predictions obtained from the WLLN. To take an example, with $N=10^{10}$ (which is already a large number for a quantum experimentalist) and $\mathcal{P}_\spadesuit=\frac{1}{2}$ we see from Eq.~\ref{err2} that the probability for the frequentist Born rule $Q_\spadesuit \simeq \frac{1}{2}\pm \varepsilon$ to hold true experimentally with an uncertainty of at most $\varepsilon\simeq 4\cdot 10^{-5}$, i.e., corresponding to  $\sigma=10$ in Eq.~\ref{err2}, is $\sim 1-10^{-23}$. Furthermore, for a `macroscopic' sample involving $N\sim 10^{24}$ quantum coins the Born rule $Q_\spadesuit \simeq \frac{1}{2}\pm 5\cdot 10^{-11}$ is obtained for $\sigma=100$ with a gigantic probability (also named confidence level) $\sim 1-10^{-2174}$. But what is the physical meaning  of an atypical event with probability like $10^{-20}$ or $10^{-2000}$? \\
 \indent Already in 1713 Bernoulli considered events with small probabilities as `morally impossible' and those with high probabilites as `morally certain'. While we don't here accept  a subjective approach of probabilities the same issue was discussed more objectively at the end of the $18^{th}$ century by d'Alembert and Buffon, and in 1843  by the philosopher and mathematician Cournot who wrote:
 \begin{quote}
\textit{ The physically impossible event is therefore the one that has infinitely small probability, and only this remark gives subtance- objective and phenomenal value- to the theory of mathematical probability.}~\citep[p.~78]{Cournot1843}
\end{quote} This `Cournot principle', as it was later named, is sometimes considered as a corner stone of modern probability theory (for a review see~\citep{Shafer2007}), although it belongs more to the applicative  and empirical side of the theory than to the abstract measure-theoretic side associated with pure mathematics and enumeration of alternatives.\\
\indent Borel, for example, considered that negligible probabilities associated with atypical events (and not only infinitessimal as believed by Cournot~\footnote{Cournot analysis was done at a time when a clear understanding of the notion of limits and infinite sets was still missing. In particular, the difference between uncountable and countable infinite sets associated with the work of Cantor, Lebesgue, Borel and many others was yet to be developped. Despite this Cournot got the good intuitions concerning the notion of physically impossible events. For instance, he identifed the infinitely small probability associated with  a particular sequence in an infinitely long Bernoulli process with the infinitely small probability associated with a point on a disk. This clearly predated the discussions made at the start of the $20^{th}$ century concerning null-measures and countable additivity (for an historical discussion see \citep{Martin1994}).}) are necessary in order to attribute a physical content to the WLLN and to define an `empirical law of randomness'~\citep[p.~54]{Borel1923}. In a short article `on universally negligeable probabilities' published in 1930~\citep{Borel1930}, Borel wrote:
\begin{quote}
\textit{The necessary conclusion is that the probabilities that can be expressed by a number smaller than $10^{-1000}$ are not only negligible in the common practice of life, but universally negligible, that is they must be treated as rigorously equal to zero in all questions concerning our Universe. }~\citep{Borel1930}
\end{quote} We stress that for Borel such negligible probabilities were both subjectively and physically negligible~\citep{Borel1939}. The analysis of such atypical events are therefore representing a central element for interpreting both subjective and objective probabilities~\footnote{As emphasized by Borel \citep{Borel1914,Borel1939,Borel1930}, we `certainly know' as a rule of thumb that when a gambler is predicting that `for sure'  the next long-run $N\sim 10^{10}$ of a Bernoulli game  will produce an atypical sequence with $Q_\spadesuit$ strongly deviating from $\mathcal{P}_\spadesuit$  (e.g., $\varepsilon\simeq 4\cdot 10^{-5}$) we can almost be certain that this gambler is lying:  the certainty being here defined, at least, with a probability like $\sim 1-10^{-23}$.} and for giving a `practical and philosophical value to probabilities'~\citep{Galavotti2019,Borel1939}. \\ 
\indent Kolmogorov (following Markov, Chuprov and many others in Russia~\citep{Shafer2006}) also included this `Cournot principle' as part of the physical axioms needed to connect the abstract probabilistic formalism to the actual world, and wrote:
\begin{quote}
\textit{if} $\mathcal{P}(A)$ \textit{is very small, one can be practically certain that when conditions} $E$ \textit{are realized only once, the event} $A$ 
\textit{would not occur at all.}~\citep[p.~4]{Kolmogorov1950}
\end{quote}
Following all these mathematicians, we are tempted to say that great probabilities, i.e., ratios like  $\frac{M_{\mathcal{H}_{\textrm{typical}}}}{2^N}\sim 1-10^{-2174}$ are for all practical needs identified with negligible deviations from the absolute certainty. In other words, atypical events, i.e., maverick histories, are completely negligible in probability. Their atypicality should  make them `physically' or `factually'  impossible.\\
\indent To be more precise on the justification and application of this Cournot's principle  and to understand the  physical meaning of Borel's `universally negligible probabilities' we must ask ourself how can we measure an `un-probability' like $10^{-1000}$? In turn this must explain what we mean by `factually impossible'. If we can't answer this question we are apparently going to  enter the infinite regress loop that is (implicitly) alluded to in Wallace quote~\citep{Wallacevideo}. However, we first note with Borel~\citep{Borel1949} that if during an experiment $(E)$ (let say a Bernoulli sequence of $N$ trials of a quantum coin tossing) an atypical event  has a tiny probability $\mathcal{P}_{\textrm{atypical}}=1/n$, with $n\gg 1$, then the probability to observe at least one time this atypical event in a super-Bernoulli sequence of $n$ trials of the experiment $(E)$ is $1-(1-1/n)^n\simeq 1-1/e\simeq 63\%$.  Such a probability is clearly not anymore negligible~\footnote{Note that by pushing the number of trials to $100 n$ we would obtain a probability $1-(1-1/n)^{100n}\simeq 1-e^{-100}\simeq 1- 10^{-43}$, i.e, a quasi-certainty~\citep{Borel1930,Borel1949,Borel1950}. We also stress that the notion of typicality and negligible probabilities advocated by Borel was popularized in his book `le hasard'\citep{Borel1914} with the illustrative example of a group of monkeys using typewritters for randomly reproducing pages of famous books. } and the event is expected to be observed (but without certainty of course since we are only considering probabilities, i.e., possibilities \`a la Laplace): This is the probabilistic meaning of the usual sentence `the atypical event as a chance of 1 over $n$ to happen'.\\ 
\indent  At that stage we still have a WLLN statement:  `a frequency is a probability  with a high probability'. Therefore, in order to avoid the infinite regress loop, we must introduce an additional element. Moreover, with Borel~\citep{Borel1949,Borel1950}, we  observe that very quickly, i.e., for $\sigma$ large enough, the number $n$ becomes extremelly large and goes far beyond the possibilities physically allowed by the present Universe. To understand what are these limits suppose for example a cosmological system (`our Universe') with $m\sim 10^{81}$ atoms and consider, as an estimation, that the number of trials physically possible for each atom is given by the ratio $T_U/T_P\sim 10^{62}$ of the typical `age' of our Universe $T_U$ to the Planck time $T_P$. We thus  get a maximal total number of trials 
$n_{\textrm{max}}\sim m\frac{T_U}{T_P}\sim 10^{143}$ allowed in our cosmological `Hubble volume' (i.e., limited by causal considerations and the specific model used). This number should be compared with $n$, or more precisely $nN\simeq n$, the whole number of elementary trials in the super-Bernoulli process we want to realize. Therefore, if we consider $\mathcal{P}_{\textrm{atypical}}:=\mathcal{P}(|Q_\spadesuit-\mathcal{P}_\spadesuit|\geq \sigma_{\textrm{max}}\delta Q_\spadesuit)=1/n_{\textrm{max}}\sim 10^{-143}$, i.e.,  the probability for an atypical event  in a Bernoulli sequence $(E)$, we obtain from Eq.~\ref{err2} (see Footnote \ref{foot1}) the largest possible value for $\sigma_{\textrm{max}}\simeq 25.5$ corresponding to $\varepsilon_{\textrm{max}} =\sigma_{\textrm{max}}\delta Q_\spadesuit\simeq \frac{12}{\sqrt{N}}$. This represents the physical limits for the application of the WLLN in our Universe~\footnote{We emphasize that the discussion presented here is limited to the Bernoulli process based on the binomial distribution. To take a known analogy,  this corresponds  to the problem of drawing  
a ball  ($N$ times) in a urn with replacement (the urn containing $M_\spadesuit$ and $M_\heartsuit$ balls respectivelly).  Alternatively, if we consider drawing of a ball without replacement we have to use the hypergeometric distribution. In this case where the urn contains a finite number of balls $M=M_{\spadesuit}+M_{\heartsuit}$ the maximum number of extractions without replacement is naturally limited by $n_{\textrm{max}}:=M$, i.e., the sample size can not be larger than the whole population. We stress, that we believe the hypergeometric distribution to be a good starting point for a clean foundation of probability. In this approach the population $M_\spadesuit/M$ constitutes a definition of the probability $\mathcal{P}_\spadesuit$ which can therefore be identified with the Laplace-Gibbs representation of the not-so fictitious probablity fluid given by Eq.~\ref{Laplacedef}.}. Beyond this value, i.e., $\sigma>\sigma_{\textrm{max}}$, we have   $\mathcal{P}_e=1/n < 10^{-143}$ which is universally negligible. \\
\indent Most importantly for us, in discussing the experimental verification of the results predicted by the probability calculus Borel wrote:
 \begin{quote}
\textit{the verification consists then to notice that the event whose probability is very low, that is negligible at the cosmic scale never happens. }~\citep[p.~12]{Borel1939}
\end{quote} This idea of an event whose probablity is sufficiently small to never occur is central for understanding and justifying the application of the algorithmic probabilistic method to the physical world. It represents the ultimate (also named strong) interpretation of Cournot's principle assumed by Borel and allows us to state that actual events are always typical. In other words, what is needed in order to break the infinite regress paradox is to eliminate atypical events from the actual world:
 \begin{quote}
\textit{Atypical events \underline{are never} happening}. 
\end{quote}
This is the true message of typicality. Clearly, when  we say that maverick alternatives are not happening we don't say that they could not happen. Better, if atypical events are not happening this is because of a choice made by the Universe itself. In a deterministic Universe for example, assuming fixed evolution laws like Newtonian or Bohmian mechanics, the initial conditions on a distribution of, let say, particles over a space-like hyper surface $\Sigma_0$ is all what is needed to fix the subsequent evolution of the system. This initial condition choice is not law-like (as the dynamical laws are) but better fact-like, i.e., contingent, since other choices were clearly possible and available in the configuration or phase space~\footnote{The different between a fact-like and law-like description is often arbitrary since many laws of nature that are once considered as fundamental  are later accepted as emergent and are derived from more fundamental hypotheses (e.g., the thermodynamic relations are now derived from statistical mechanics). }. In purely stochastic theories, like GRW~\citep{GRW}, the ontology is about a cloud of events (e.g., a Bernoulli sequence) distributed over space-time and there is no law for the individuals only for the collective. Moreover, contrarily to the claims made by von Mises~\citep{vonMises1957} this collective has not to be infinite. The WLLN associated with the strong Cournot principle leads to a statistical Born rule emerging at the collective level for $N$ large but finite. In this approach atypical histories with low probablities are simply not happening. In a certain sense, we could say that this is a conspiracy of nature albeit a typical one, i.e., a conspiracy that looks so natural to us that we dont' see it as extraordinary (at the difference of superdeterminism or retrocausality).    \\
\indent A misguided (but interesting) objection to this idea of an universally small probability regime (and therefore of atypicality) is that, after all, every branches $h$ of the Bernoulli tree have a small probability (e.g., $\mathcal{P}_h= 1/2^N$ for the case $\mathcal{P}_{\spadesuit/\heartsuit}= 1/2)$. Therefore, if $N$ is large enough we have always $\mathcal{P}_h\ll 10^{-143}$ that is universally negligible, and consequently the event $h$ should (according to a naive reading of Cournot principle) not occur. But since this is true for any $h$  none of the history $h$ should be allowed to occur and we have thus a paradox (this paradox was discussed by Borel in \citep[p.~15]{Borel1939}). The response to this paradox, is that when we speak of facts we are actually considering a Universe containing a machine or device programmed to find a specific event: for example the device could be programmed for finding the single history  $h_1:=[\spadesuit,\spadesuit,...,\spadesuit,\spadesuit]\in \mathcal{H}_{\textrm{atypical}}$ or alternatively the machine could be programmed for finding  $h_2:=[\spadesuit,\heartsuit,...,\spadesuit,\heartsuit]\in \mathcal{H}_{\textrm{typical}}$.  The rule here, is that the only way for the machine to check if the outcome will actually occur is to try again and again (assuming the machine has no other information about the system),  and the notion of atypicality advocated here says that this event with universally negligible probability  will never happen in our Universe: this is a postulate.  Now, to say that a machine or `observer' programmed to find $h_1$ (or $h_2$) will not succeed is certainly different from telling that no machine at all will find $h_1$ (or $h_2$). The central difference is that in the former case the machine was programmed to find $h_1$  (or $h_2$) and in the latter not. Moreover, in the real world we, as devices, are not interested into such fine grained information  but rather on the frequencies $Q^h_{\spadesuit/\heartsuit}$. It is with these observables $Q^h_{\spadesuit/\heartsuit}$ that probabiliies $\mathcal{P}(N_\spadesuit)$ are defined and the WLLN is obtained (see Eqs.~\ref{laplace},\ref{err}). The notion of typicality and universally negligible probabilites used by Borel have therefore acquired a factual and physical meaning through the existence of typical observers and the non existence of atypical ones. \\
\indent This reading of typicality looks probably extreme, or even odd, to many but we believe it is the only one which is devoid of contradiction and circularity. Indeed, typicality is a mere `geometrical' statement about the size of a set $\mathcal{H}_{\textrm{typical}}$ relatively to $\mathcal{H}$  assuming a measure on these sets. Mathematically, it is just a characterization of a set.  Yet, there is nothing  in this definition which can force the system in the typical set to be actual or not (even if the typical set is overwhelmingly dominating the whole ensemble at a measure-theoretic level).  Therefore, this dilemma requires a different axiom, as we suggested, in order to  break the infinite regress paradox and to avoid to call typicality a pure tautology.\\
\indent  Therefore, the suggestion `atypical events \underline{are not} happening' is mathematically to reverse the order of the axioms in probability theory for all practical purposes.  Here, we start with actual  frequencies $Q_\spadesuit$, $Q_\heartsuit$ recorded during a Bernoulli sequence with $N$ independent trials~\footnote{The notion of independence here means that we reproduce the same external conditions for the experiments and that the chain rule for independent probabilities apply. This is of course only a definition that is justified only by the physical consequences that we can get from it.  Here we assume that nature satisfies this rule.} and we consider $\mathcal{P}_\spadesuit$, $\mathcal{P}_\spadesuit$ as estimators associated with a fictititous fluid satisfying the probability axiomatics (in particular the so called chain rule for independent trials that is fundamental for deriving the WLLN).  The aim of these estimators is to reproduce acurately the recorded data $Q_\spadesuit$, $Q_\heartsuit$. And the best fit is, following the WLLN, obtained when $Q_\spadesuit\simeq \mathcal{P}_\spadesuit$, $Q_\heartsuit\simeq\mathcal{P}_\heartsuit$. Naturally, this inversion between the leading roles of frequencies and probabilities is only a practical rule of thumb, i.e., for an experimentalist interested into frequencies. Moreover, objective probabilities are fundamentally more important since they reveal an attractor for frequencies using the WLLN. Therefore, probabilities, like temperature or pressure, are objective properties of the whole system (or population) that are only probed and approched with samples of size $N$.  Importantly, our principle or method is valid for $N\gg1$ (i.e., in order to apply  Stirling's approximation needed for obtaining Eqs.~\ref{laplace},\ref{err2}).  We dont require the system to be actually or hypothetically infinite (unlike von Mises with his frequentist theory of probability \citep{vonMises1957}). Additionally, the hypotheses leading to the WLLN are such that the equalities $Q_\spadesuit\simeq \mathcal{P}_\spadesuit$, $Q_\heartsuit\simeq\mathcal{P}_\heartsuit$ are not to be strict.  The formalism allows for fluctuations so that for a   recorded value $Q_\spadesuit$ we can define a confidence interval for $\mathcal{P}_\spadesuit$. That is, for $N\gg 1$ we have approximately~\footnote{The confidence interval for $\mathcal{P}_\spadesuit$ is obtained by inversion from the rigorous confidence interval for $Q_\spadesuit$, i.e, $|\mathcal{P}_\spadesuit-Q_\spadesuit|
\leq \sigma_{\textrm{max}}Q_\spadesuit$. The rigorous inversion leads to: 
\begin{eqnarray}
|\mathcal{P}_\spadesuit-\frac{Q_\spadesuit+\frac{\sigma_{\textrm{max}}^2}{2N}}{1+\frac{\sigma_{\textrm{max}}^2}{N}}|
\leq \sigma_{\textrm{max}}\frac{\sqrt{(Q_\spadesuit(1-Q_\spadesuit)+\frac{\sigma_{\textrm{max}}^2}{4N})}}{\sqrt{N}(1+\frac{\sigma_{\textrm{max}}^2}{N})}
\end{eqnarray}  that reduces to Eq.~\ref{eqconf} for $N\gg 1$.  } 
\begin{eqnarray}
|\mathcal{P}_\spadesuit-Q_\spadesuit|
\leq \sigma_{\textrm{max}}\frac{\sqrt{(Q_\spadesuit(1-Q_\spadesuit))}}{\sqrt{N}}\label{eqconf}
\end{eqnarray} with a confidence level $\textrm{erf}(\frac{\sigma_{\textrm{max}}}{\sqrt{2}})\simeq 1-10^{-143}$ identified with certainty.\\
\indent Before to conclude this section we remind that in physics the importance of typicality was already stressed long ago by Boltzmann~\citep{Boltzmann1877,Boltzmann1898,Boltzmann1896} in his research for a clean foundation of statistical mechanics. For instance, replying to a well-known criticism made by Zermelo on the consistency of the probabilistic approach, Boltzmann wrote:
\begin{quote}
\textit{I assert on the contrary that by far the largest number of possible states are ``Maxwellian'' and that the number that deviate from the Maxwellian states is vanishingly small.  }~\citep[p.~395]{Brush2003}
\end{quote} In the last decades, Boltzmann's idea that discussions about typicality, i.e., about very large and very small probabilities, must play a central role for interpretating statistical mechanics and also quantum mechanics (e.~g., in the de Broglie Bohm interpretation~\citep{Durr1992}) as evolved into a `typicality school'~\footnote{For recent philosophical discussions about some controversial features of typicality and probability in the context of statistical, quantum  mechanics, and cosmology see \citep{Maudlin2007,Volchan2007,Werndl2013, Lazarovici2015,Barrett2017,Wilhelm2019,Allori2020,Maudlin2020,Hubert2021,Valentini2020}. }, famously advocated by Lebowitz~\citep{Lebowitz1993}, Goldstein~\citep{Goldstein2012}, and Penrose~\citep{Penrose2010}. We emphasize that sometimes for advocates of the typicality school, typicality is considered as more fundamental than probability (i.e., too often assumed as a purely epistemic concept), and an opposition is made between Boltzmann and Gibbs. Here, we dont subscribe to this radical view: typicality should not be thought as an alternative to probability, better typicality, i.e., through the strong Cournot principle, provides a practical and algorithmic way to use probabilistic deductions (i.e., it makes the calculus of probabilities safe of contradiction). Furthermore, for advocate of the typicality school, the typicality measure (e.g., the Liouville/Lebesgue measure in statistical mechanics or the equivariant $|\Psi|^2$ measure in Bohmian mechanics) is supposed to have a natural, preferred physical meaning justifying his unambiguous use.  Once again, we  disagree:   the choice of a typicality (i.e., probability) measure is actually guided by experiments and should not be (only) grounded on mathematical symmetry or elegance (even if we agree that equivariance is a good characterization of the actual state of our Universe obeying Born's rule). 
\section{Why the many-worlds interpretation is typically dead} \label{sec4}       
\indent Assuming the previous interpretation of typicality and universally negligible probabilities we can go back to the MWI proposed by Everett~\citep{Everett1957,Barrett2012,DeWitt1973}. As we explained, the MWI preserves the purely unitary evolution  and attempts to give an ontic value to the quantum state $|\Psi\rangle_t$ driven at every time  $t$ by Schr\"odinger's equations $i\hbar \partial_t|\Psi\rangle_t=H|\Psi\rangle_t$. Following this interpretation, if we  consider a quantum Bernoulli process, like the one described in section \ref{sec2}, we must assume that all branches $h$ in the history tree $\sum_h a^{N^h_\spadesuit} b^{N^h_\heartsuit}|\mathcal{M}_0,\mathcal{A}\textrm{ur\'elien}_h\rangle_{t_f}$, i.e., associated with Eq.~\ref{eq3} and depicted in Fig.~\ref{figure1}, must be considered on an equal footing. There is no collapse in the MWI, and, unlike in the Copenhagen interpretation, the wave-function $|\Psi\rangle_t$ can not just be used as a book-keeping for quantum alternatives before the actualization of only one history $h_0$.\\
\indent  In this context, it is not very difficult to see that the mere philosophy of the MWI directly conflicts with the notion of typicality discussed in section~\ref{sec3}. Indeed, if all branches should be preserved in the full unitary evolution $U_t$,  then $U_t$ necessarily includes so-called maverick or atypical branches which, following our Cournot-Borel principle `atypical events \underline{are not} happening', must be forbidden to occur. It is here central to compare the MWI with `single-world' interpretations (like the Copenhagen or Bohmian interpretation)  which assume the different alternatives $h$ as virtually existing in a configuration or phase space.  In the end, for these various single-world frameworks there is only one reality for the quantum system and only one $h$ is actualized (whatever `actual' meant in these approaches). The fact that this actualization occurs, in agreement with Cournot's principle, in the typical set  $\mathcal{H}_{\textrm{typical}}$ satisfying Born's rule is not contradictory and can safely be made an integral part of the ontology. That is clearly not the case in the MWI and maverick branches can not just be elimitated without changing the ontology (i.e., by modifying the dynamics or adding some new ideas as suggested in the introduction).\\
\indent Interestingly enough, this issue has always been known, even though, perhaps not fully appreciated by all advocates of the MWI. One of the earliest discussion of this topic was probably done by DeWitt who, after reminding that `all the worlds are there, even those in which everything goes wrong and all the statistical laws break down', wrote:
\begin{quote}
\textit{We can perhaps argue that in those branches in which the universe makes a habit of misbehaving on this way, life fails to evolve; so no intelligent automata are around  to be amazed by it.}~\cite{DeWitt1971}
\end{quote} 
In other words, it is claimed that the maverick branches are not existing because they violate some entropic and anthropic rules which would otherwhise conflict with our own  physical existence.  This is somehow introduced by \emph{fiat} as a contingent fact. Moreover, DeWitt even goes to suggest that 
\begin{quote}
\textit{It is also possible that maverick worlds are simply absent from the grand superposition.}~\cite{DeWitt1971}
\end{quote} DeWitt subsequently speculated~\footnote{In the recent years such proposals have been considered seriously by Hanson~\citep{Hanson2003} who introduced `mangled' worlds. This speculative theory could perhaps be formalized by using the strong Cournot principle to modify amplitudes $\mathcal{A}_h=a^{N^h_\spadesuit} b^{N^h_\heartsuit}$ in  Eq.~\ref{eq3}.  More precisely,  the new amplitudes (solutions of a future non-linear quantum theory to be developped) would thus read: \begin{eqnarray}
\mathcal{A'}_h=a^{N^h_\spadesuit} b^{N^h_\heartsuit}\textrm{Rect}\left(\frac{Q_\spadesuit-\mathcal{P}_\spadesuit}{2\sigma_{\textrm{max}}\delta Q_\spadesuit}\right)
\end{eqnarray} where $y=\textrm{Rect}(x)$ is a normalized boxcar function such that $y=1$ if $|x|\leq \frac{1}{2}$ and $y=0$ otherwhise. All branches outside the confidence interval given by Eq.~\ref{eqconf} are thus eliminated from the `grand superposition'. For a different but related  idea see~\citep{Buniy2006}. An other approach would be to assume the many de Broglie Bohm ontology~\citep{Tipler2014,Bostrom2014,Sebens2014} and assumes that there is only a finite number of world-lines associated with the population of Universes.  This population is distributed according to Born's rule and during a long  Bernoulli process one could imagine that some atypical branches would be ultimately empty because the population of world-lines is not large enough to fill every branches (this is perhaps the meaning given by DeWitt in \citep{DeWitt1971}).} about the possibility of a finite Universe containing only a finite number of branches~\cite{DeWitt1971}. Here we are not going to discuss nonlinear extensions of the MWI but these are serious possibilities to be considered. \\
\indent  Yet, the importance of maverick worlds or branches has been recently underlined by McQueen and Vaidman~\citep{McQueen2019} in a response to Albert~\citep{Albert2015}. In complete agreement with Dewitt (at least on that issue) they indeed wrote:    
\begin{quote}
\textit{But now there are two responses: (i) once we start talking about worlds with different natural processes, we begin talking about worlds that may not support life and therefore measurement outcomes; and (ii) even if such worlds do support life, we find it very implausible that it would be possible to construct a reasonable theory of natural phenomena in such worlds.}~\cite{McQueen2019}
\end{quote} The explanation of McQueen and Vaidman is motivated by the critical analysis made by Albert~\citep{Albert2015} concerning the absence (in the MWI) of empirical confirmations of the statistical Born's rule $Q_\alpha \sim\mathcal{P}_\alpha$ for atypical histories $h\in\mathcal{H}_{\textrm{atypical}}$. Albert wrote: 
\begin{quote}
\textit{Why (for example) should it come as a surprise on a picture like this, to see what we would ordinarily consider a low-probability string of experimental results? Why should such a result cast any doubt on the truth of this theory (as it does, in fact, cast doubt on quantum mechanics)?}~\citep[p.~162]{Albert2015} 
\end{quote} We completely agree with Albert, and we actually think that this is a fatal objection against the MWI. Interestingly, McQueen and Vaidman~\citep{McQueen2019} try an answer by suggesting a difference between confirmation and deduction of a probability rule. They suggest that the probability rule must be deduced from the theory itself, i.e., the MWI, including all branches $h$, but that only the  experimental records of observers belonging to the typical set $\mathcal{H}_{\textrm{typical}}$ are physically significant for us as empirical confirmation (i.e., assuming we are typical observers). The previous quote of McQueen and Vaidman concerning  the difficulty or impossibility of confirmation for an observer belonging to  $\mathcal{H}_{\textrm{atypical}}$ is linked to this suggestion of DeWitt of prohibiting atypical worlds, i.e., by suggesting that life or sophisticated creatures able to records data are simply absent in those maverick branches. Furthermore,  note that McQueen and Vaidman~\citep{McQueen2019} assume that the situation is not better no worst that it is in other statistical and quantum interpretations. Replying to Albert they wrote: 
\begin{quote}
\textit{Consequently, we can ask Albert's question: Why should it come as a surprise, on a collapse theory, to see what we would ordinarily consider a low-probability string of experimental results? After all, such a string is logically consistent with a collapse theory. The answer is the same as in the MWI: there is no reason for why one should take such a string to refute collapse theories if a much broader set of data (including e.g. the reason for the sky's colour) supports the Born rule.}~\cite{McQueen2019}
\end{quote}
But assuming Cournot's  principle we believe that the previous analysis is inappropriate: it is definitely not a question of being `logically consistent'. According to strong-Cournot's principle  atypical strings don't occur not because of a dynamical law but because of a contingent choice made on the initial conditions of the Universe (i.e., in a deterministic case) or because of a choice done in the whole tapestry of the Universe in a long-run sequence (i.e., in a stochastic case). This must be assumed as a fundamental property of the Universe and this even if logically speaking the laws of physics don't prohibit maverick worlds. It is interesting to observe that the paradoxical conclusions obtained by McQueen and Vaidman are the same as could be made by somebody objecting against the self-consistency of the probability calculus, i.e., by ignoring the strong Cournot principle. This inappropriate objection states that the observation of an event with very low probability would not invalidate the principle of the probability calculus since this theory only deals with probabilities  (that are by definition associated with the absence of certainty), whereas the non observation of such an event would not prove the theory either. In a sense the theory of probablity would be irrefutable or unfalsifiable and, as a consequence, non-scientific! This objection misses the point that one must necessarily include Cournot's principle in any application of the whole probability method to the real physical world.  This is actually equivalent to assuming that all observed events are typical, and therefore, that atypical or maverick events never occur. Once again, this strong Cournot principle makes perfectly sense in a single-world picture if the ontology can afford this constraint (e.g., by playing with the initial conditions like in Bohmian mechanics). This is not true in the MWI where unitarity requires all branches, even very atypical ones, to exist. Therefore, the analysis of DeWitt and McQueen-Vaidman are untanable and indefendable if we stick to the framework of the pure unitary MWI~\footnote{A recent article by Saunders reproduces the same error. He wrote `Denizens of anomalous branches, or of anomalous stretches of history in one-world theories, will be misled by the observed statistics of measurements. They will conclude that quantum mechanics (or at least the Born rule) is false. But they will simply be unlucky'~\citep{Saunders2021}. This  inappropriate notion of `bad luck' conflicting with Cournot's principle is also advocated by Tappenden~\citep{Tappenden2019} in his criticism of Adlam work~\citep{Adlam2014}.}.  \\
\indent Going back to this difference between confirmation and derivation of Born's rule advocated by McQueen and Vaidman, we already stressed in the introduction that many authors in the past have correctly emphasized the circularity and difficulties of all the purported proofs of Born's postulate in the MWI~\citep{Ballentine1973,Benioff1978,Stein1984,Kent1990,Squires1990,Putnam2005,Hemmo2007,Adlam2014,Barrett2017}. Putnam~\citep{Putnam2005}, for instance, like Hemmo and Pitowsky\citep{Hemmo2007}, have questionned why a simple branch counting should not be favored instead  of the $|\Psi|^2$-rule. More recently, Bricmont~\citep{Bricmont2016} has been very clear about it. Commenting about what will occur in a asymmetric  situation where $\mathcal{P}_\spadesuit=3/4$ he wrote: 
\begin{quote}
\textit{But then the same use of the law of large numbers leads to the conclusion that, in the vast majority of worlds, the quantum predictions will not be observed, since our descendants will still see $N/2$ splittings where the cats end up alive and $N/2$ splittings where they end up dead, instead of the (3/4,1/4) frequencies.}~\citep[p.~203]{Bricmont2016}
\end{quote}  Despite many attempts made by advocates of the MWI over the years, we believe that this Bricmont objection about the simple branch-counting constitutes a valid objection against the standard MWI. Moreover, the situation is even worse than that:  The question is not to know if the simple branch-counting could be a natural alternative to Born's rule~\footnote{Note, that the motivation for the simple branch-counting procedure considered here should not be confused with a naive perspective that we could call the `god-eye perspective'  for which the observer metaphysically located outside the Universe count the branches.  Instead,  here we are correctly discussing the inner perspective advocated by Everett where the observer is an integral part of the unitary evolution in the Bernoulli tree.   Such an  observer is not aware of the existence of the other branches.}. Better, the question is to understand  whether or not any probability rule can be chosen unambigously in the current framework of the MWI. We believe that the answer is not, and that the MWI can not survive to this criticism without important amendments. In a remarkable paper Adlam~\citep{Adlam2014} has been  very lucid about that issue and wrote
\begin{quote}
\textit{Moreover, the Everett approach entails not only that mod-squared amplitudes cannot play this role but, but also that nothing else in the theory can play it either: since the theory does not single out any one sequence of outcomes, no entity defined within that theory can be responsible for our having seen the particular sequence of results that we have, and it follows  that in an Everettian universe we would not be able to establish reference to  the theoretical entities required by the Everettian theory.}~\citep[p.~26]{Adlam2014}
\end{quote}  
Once again, we fully agree with this diagnostic:  Both the Bricmont and Adlam analysis drive the final nails into the coffin of the old MWI. Moreover, here we want to play the role of the conorer and,  based on our previous analysis of typicality, we  actually  provide the autopsy report of the MWI.\\
\indent Remind first, Borel dictum `Atypical events \underline{are never} happening', and the consequence this message has on the pragmatic use of probabilities $\mathcal{P}_\alpha$ as extimators for frequencies $Q_\alpha$. Remind also, that in quantum mechanics the Bernoulli process represents a kind of fractal tree skectched in Figure.~\ref{figure1}.                        This tree is before all a topological structure and if we want to define a probability measure on this tree we must define a kind of fundamental metric for weighting the different branches:  this what is named defining the theory by McQueen and Vaidman~\cite{McQueen2019}.\\ 
\indent However, for any  branches $h$ on the tree  we can define a frequency $Q^h_\spadesuit$ and by applying Eq.~\ref{eqconf} for a number of trials  $N\gg 1$ we can \underline{always} estimate  $\mathcal{P}_\spadesuit\sim Q^h_\spadesuit$ by using the parameter $\sigma_{\textrm{max}}$ associated with universally negligible probabilities \`a la Cournot-Borel. And note again, that since the MWI is rigorously preserving unitarity, this is actually valid for every branches $h$, i.e., every sequences of outcomes.    
Therefore, once we assume the existence of all branches we have no other choice than to accept that the theory contains \underline{all} the probability laws, i.e., including the Born rule but also the simple branch-counting and anything else we want only constrained by the law $\mathcal{P}_\spadesuit+\mathcal{P}_\heartsuit=1$. In other words, in the MWI in order to fit $Q^h_\spadesuit$, we can arbitrarily change the metric $\mathcal{P}_\spadesuit, \mathcal{P}_\heartsuit$ and subsequently the domain of typicality on the branches of the fractal Bernoulli-tree.  And note once again, that this peculiar situation is specific of the MWI.  In all accepted single-world quantum interpretations with collapse (like GRW) or not (like the PWI) there is no problem because we can postulate Born's law as an external rule for univocally reproducing empirical statistics.  As a matter of fact, the MWI, as it is built only from unitarity, can just not be true since it can not unambigously impose Born's rule.\\
\indent   The previous conclusion has its cost. If every frequencies are allowed in the MWI then maverick worlds  that looks very atypical using Born's rule have necessarily their place in the fractal Bernoulli-tree. Consider for example the case where $a=b=1/\sqrt{2}$ in Eq.~\ref{eq1}, i.e., where Born's rule apriori imposes  $\mathcal{P}_\spadesuit=\mathcal{P}_\heartsuit=\frac{1}{2}$.  The typicality reasoning based on the WLLN imposes thus $Q_\spadesuit\simeq \frac{1}{2}$ as it should be.   However,  atypical branches such as  $Q^h_\spadesuit\simeq 0$ or $Q^h_\spadesuit\simeq 1$ can not be prohibited in the MWI. Theses worlds will be associated with their own regimes of typicality where  $\mathcal{P}_\spadesuit\simeq 0$ or $\mathcal{P}_\spadesuit\simeq 1$. And this is the only way to make sense at the same time of the MWI and of Cournot's principle.  In other words, this is the only solution for building a theory  that preserves unitarity together with the self-consistency of the probability calculus. Moreover, this theory, i.e., the bare MWI, clearly disagrees  with experiments imposing Born's rule: Consequently, the bare MWI must be rejected!~\footnote{We stress that this rejection constitutes a way for solving the problem of the quantum suicid or `quantum russian roulette' popularized by Tegmark~\citep{Tegmark1998} and relying on the existence of a extremelly atypical branch where the player is always surviving and wining:   This atypical branch is on the long run violating Cournot's principle and thus Born's rule...   The impossibility of such a crazy situation is clearly demonstrative.}  \\    
\indent At this step of the analysis, many proponents of the MWI should probably getup on the stage to oppose this conclusion.  After all, quantum mechanics has a preferred way to weight branches.  Born's rule $\mathcal{P}=|\psi|^2$ is not arbitrary: it is imposed by quantum mechanics itself. Indeed (would oppose a typical many-worlder), we have several ways to  justify and introduce Born's rule: (i) We can obtain it from   Schr\"odinger's equation $i\hbar \partial_t|\Psi\rangle_t=H|\Psi\rangle_t$ as a consequence of the local law of conservation $-\partial_t|\psi|^2=\nabla \cdot J_\psi$ where $J_\Psi$ is a current.  This is obtained rigorously from Noether's theorem under phase (gauge) invariance. (ii) We have Gleason's and Everett's reasoning motivating  the choice of an additive measure $f(|a|^2)+f(|b|^2)=f(|a|^2+|b|^2)$ \citep{Gleason1957,Everett1957}. (iii) we have `equivariance' that also motivates the choice $\mathcal{P}=|\psi|^2$ from current conservation (this is often used by Bohmians~\citep{Goldstein2007,Drezet2017}). (iv) We have  also more Everettian-like derivations based on Deutsch-Wallace decision theory~\citep{Deutsch1999,Wallace2007,Saunders2005,Saunders2008,Wallace2012} or `envariance' \`a la Zurek \citep{Zurek2003,Zurek2005,Barnum2003,Sebens2016}. (v) Finally, we have Vaidman derivations~\footnote{We stress that for Vaidman there is no real derivation of the Born's  rule that do not assume additional axioms~\citep{Vaidman2012,Vaidman2020}. Moreover, following our analysis of typicality and Cournot's principle, Vaidman deductions motivated by locality can only work in those branches where Born's rule hold. This is thus circular and inappropriate for `maverick branches'. The same is true for decision theoretic derivations \`a la Deutsch Saunders Wallace~\citep{Deutsch1999,Saunders2008,Wallace2012} where Born's `epistemic' probability, i.e., defined as a subjective credence, is assigned to every branches including Maverick ones where the rule  contradicts bare facts (that is in conflict with the Lewis principal principle).} based on locality, symmetries and equiprobability \citep{Vaidman2012,McQueen2019,Vaidman2020}. All these derivations, in addition to unitarity, require some `natural' hypotheses that are motivated  by empirical facts (like locality  and equiprobabilities based on Keynes-Laplace's indifference principle) or elegant symmetries (like equivariance and envariance).\\
\indent However,  despite naturalness  or elegance these results conflicts with our previous conclusion concerning the  MWI and typicality. It is not here our aim to review all the previous purported proofs of Born's rule in order to find the source of the disagreement. Instead, we want to emphasize the role of a old but often neglected result~\footnote{Over the years, this result has been nevertheless discussed by some advocates of the PWI (especially Bohm and Vigier~\citep{Vigier1954}, Bohm and Hiley~\citep{BohmHiley}, Valentini~\citep{Valentini1992,Valentini2020} including the present author~\citep{Drezet2017}). For a complete review of different probabilistic Bohmian interpretations see \citep{Callender2007}. } obtained by de Broglie~\citep{Broglie1928} and Bohm~\citep{Bohm1953} in the context of the pilot wave mechanics.  Consider once again the local conservation law  $-\partial_t|\psi|^2=\nabla\cdot J_\psi$  deduced from Schr\"odinger equation.  In the pilot wave interpretation  we introduce the velocity field $v_\psi(x,t)=\frac{J_\psi(x,t)}{|\psi|^2}:=\frac{dx(t)}{dt}$ (associated with the motion of the beables with coordinates $x(t)$ in the configuration space) and the conservation law reads \begin{eqnarray}
 -\partial_t|\psi|^2=\nabla \cdot (v_\psi|\psi|^2).
\end{eqnarray}  
Moreover, de Broglie and Bohm suggested that the most general probability distribution $\rho(x,t)= f(x,t)|\psi|^2$ carried by the same velocity field $v_\psi(x,t)$ must also satisfy the conservation \begin{eqnarray}
 -\partial_t\rho=\nabla \cdot (v_\psi\rho).
\end{eqnarray} comparing the two laws of conservation for $|\psi|^2$ and $\rho$ we immediately deduce the Lagrange derivative
\begin{eqnarray}
 (\partial_t+v_\psi\cdot\nabla) f(x,t):=\frac{d}{dt}f(x(t),t)=\frac{d}{dt}f_t=0
\end{eqnarray} showing that the function $f$ is transported as a whole along a trajectory $x(t)$. This fundamental results is the counterpart of Liouville theorem's $\frac{d}{dt}\eta(x(t),p(t),t)=0$ in statistical mechanics where $\eta(x,p,t)$ is the probability density in the phase space.  To enhance this analogy, the  configuration space measure must be written $\delta\gamma(x,t)=|\psi(x,t)|^2dx$ in the configuration space, i.e.,  as the counterpart of the Lebesgue-Liouville measure  $\delta\Gamma(q,p,t)=dqdp$ in the phase space~\citep{Drezet2016}. In both cases, we have $\frac{d}{dt}\delta\Gamma_t=0=\frac{d}{dt}\delta\gamma_t$ meaning that the measure is preserved during the time evolution. Note that this result is very robust and can be obtained with all known Hamiltonians used in quantum mechanics. Furthermore, the theorem doesn't require us to subscribe to the PWI:   the velocity fluid can be seen as an hydrodynamic representation of the Schr\"odinger equation and is ontologically neutral.  Importantly, the probability $d\mathcal{P}=\rho dx$ to find the beables at point $x,t$ in the infinitessimal volume $dx$ is also written $d\mathcal{P}=fd\gamma$ where $f$ plays the role of the genuine probability density with  respect to the $\gamma-$measure. Like for the Liouville theorem the relation $\frac{d}{dt}f_t=0$ shows that once we define the density $f_0$ at a time $t_0$ we know its value at any time $t$ along the path. Therefore, if we postulate, like de Broglie did in 1928, $f_0=1$ $\forall x$ this will also be true at any time $t$, and the Born rule $\rho(x,t)=|\psi(x,t)|^2$ will be satisfied. Yet, we are not obliged to do so   and consequently  the condition $f=1$, i.e., Born's rule, is not mandatory.\\
\indent It is at that step that one must add other principles to justify $f=1$.  This can be done in the PWI by postulating some distributions over the initial conditions and we can try, like Bohm and others \citep{Bohm1953,Vigier1954,Valentini1992,Valentini2020,Drezet2016}, to justify the robustness of the condition $f\simeq 1$ as a kind of attractor under a chaotic dynamic.    However, in the MWI such studies are not relevant since the coordinates $x(t)$ are not recognized as beables of the Everettian approach (unlike in the many de Broglie-Bohm worlds~\citep{Tipler2014,Bostrom2014,Sebens2014}). The problem is that there is no place in the theory for introducing an additionnal postulate for $f=1$ without pervading the original goal of Everett. Therefore the diagnostic is irrefutable:  in order to save the MWI one must modify or amend the theory (e.g., by adding a  Born-Vaidman principle  physically justified).\\
\indent Going back to the quantum Bernoulli process studied in this article we will now detail this diagnostic and show that the wave-function $|\Psi_1\rangle=a|\spadesuit\rangle+b|\heartsuit\rangle$ allows us to discuss the role of the $f$-distribution in the MWI. To fix the idea consider that $\spadesuit/\heartsuit$ are the two $z$-spin states  of a single (non-relativistic) electron described by the wave-function
\begin{eqnarray}
|\Psi_1\rangle_t=\int d^3\mathbf{x} [\psi_\spadesuit(\mathbf{x},t)|\spadesuit\rangle+\psi_\heartsuit(\mathbf{x},t)|\heartsuit\rangle]
\end{eqnarray}   with $\psi_\spadesuit,\psi_\heartsuit$ the two components of the electron-spinor $\phi$ and $\mathbf{x}\in \mathbb{R}^3$ the spatial electron coordinates. Within the Bohm-Schiller-Tiomno theory~\citep{Bohm1955}, that expresses the PWI in the context of the Pauli equation  for the non-relativistic electron with  spin, we deduce the conservation law $-\partial_t\rho_\psi=\boldsymbol{\nabla}\cdot \mathbf{J}_\psi$ where $\rho_\psi=\phi^\dagger\phi:=|\psi_\spadesuit|^2+|\psi_\heartsuit|^2$ is the generalization of Born's density  of probability (i.e., the equivariant distribution) in presence of continuous configurations space variables $\mathbf{x}$ and discrete spin variables $\spadesuit/\heartsuit$. Writing   the current $\mathbf{J}_\psi=\rho_\psi\mathbf{v}_\psi$ we obtain the hydrodynamic velocity $\mathbf{v}_\psi(\mathbf{x},t):=\frac{d\mathbf{x}(t)}{dt}$ of the de Broglie-Bohm fluid. For the present purpose the exact expression of the  $\mathbf{v}_\psi$ field is not relevant ~\footnote{The general expression for $\mathbf{v}_\psi$ is $\mathbf{v}_\psi=\frac{|\psi_\spadesuit|^2\boldsymbol{\nabla}S_\spadesuit/m+|\psi_\heartsuit|^2\boldsymbol{\nabla}S_\heartsuit/m}{|\psi_\spadesuit|^2+|\psi_\heartsuit|^2}-\frac{e}{m}\mathbf{A}$ where $e$ and $m$ are the electric charge and mass of the electron respectively, and $\mathbf{A}(\mathbf{x},t)$ is an external magnetic vector potential acting on the fluid~\citep{Bohm1955}. The current of the Bohm-Schiller-Tiomno theory can be modified in order to agree with the non-relativistic limit of Dirac's equation  by adding to $\mathbf{J}_\psi$ a spin-current term $\mathbf{J}_\psi^{(S)}=\frac{\boldsymbol{\nabla}\times(\phi^\dagger\boldsymbol{\sigma}\phi)}{2m}$\citep{BohmHiley}. }. What is instead relevant is to understand that an initial localized single-electron wavepacket  $|\Psi_1\rangle_{t_0}$ defined at time $t_0$ (i.e., with  $\psi_\spadesuit(\mathbf{x},t_0)=aF(\mathbf{x})$, $\psi_\heartsuit(\mathbf{x},t)=bF(\mathbf{x})$ and $F(\mathbf{x})$ is a narrow wave function picked on $\mathbf{x}=0$)  interacts with the magnetic field of a Stern-Gerlach apparatus and subsequently splits at time $t_f$ into two secondary and non overlapping wavepackets  $\psi_\spadesuit(\mathbf{x},t_f)=aF(\mathbf{x}+\mathbf{d}/2)$, $\psi_\heartsuit(\mathbf{x},t_f)=bF(\mathbf{x}-\mathbf{d}/2)$ (i.e., assuming that the separation $\mathbf{d}$ between the two wavepackets is larger than their width).\\ \indent If $F(\mathbf{x})$ is normalized the total $\gamma$-measure of the initial wavepacket defined as $\gamma_{t_0}=\int d^3\mathbf{x}\phi^\dagger(\mathbf{x},t_0)\phi(\mathbf{x},t_0)$ reads $\gamma_{t_0}=(|a|^2+|b|^2)\int d^3\mathbf{x}|F(\mathbf{x})|^2=|a|^2+|b|^2=1$. After the splitting the two wavepackets are associated with measures $\gamma^{(\spadesuit)}_{t_f}=|a|^2\int d^3\mathbf{x}|F(\mathbf{x}+\mathbf{d}/2)|^2=|a|^2$ and  $\gamma^{(\heartsuit)}_{t_f}=|b|^2\int d^3\mathbf{x}|F(\mathbf{x}-\mathbf{d}/2)|^2=|b|^2$ respectively, and we have naturally conservation of the total measure, i.e., $\gamma_{t_0}=\gamma^{(\spadesuit)}_{t_f}+\gamma^{(\heartsuit)}_{t_f}=1$. Moreover, from the existence of the $\mathbf{v}_\psi$-field we deduce that a fraction $\gamma^{(\spadesuit)}_{t_0}=|a|^2$ (respectively $\gamma^{(\heartsuit)}_{t_0}=|b|^2$) of the initial $\gamma$-fluid is continuously evolving to transform into the domain $\gamma^{(\spadesuit)}_{t_f}=|a|^2$ (respectively $\gamma^{(\heartsuit)}_{t_f}=|b|^2$) of the final $\gamma$-fluid.   \\
\indent Moreover, this fluid is for the moment just a mathematical measure and to transform it into a probability fluid \`a la Laplace-Gibbs we must introduce the $f$-distribution. For the present purpose it is enough to use a coarse-grained description assuming that at time $t_f$  after the splitting $f$ is constant over the finite supports of the two wave packets $\psi_\spadesuit(\mathbf{x},t_f)$, $\psi_\heartsuit(\mathbf{x},t_f)$. More precisely, we introduce $f_\spadesuit$ and $f_\heartsuit$ such that 
\begin{eqnarray}
\mathcal{P}_\spadesuit(t_f)=f_\spadesuit\gamma^{(\spadesuit)}_{t_f}=f_\spadesuit |a|^2,
\mathcal{P}_\heartsuit(t_f)=f_\heartsuit\gamma^{(\heartsuit)}_{t_f}=f_\heartsuit |b|^2,
\end{eqnarray}  with $f_\spadesuit |a|^2+f_\heartsuit |b|^2=1$ (i.e., $f_\heartsuit=\frac{1-f_\spadesuit |a|^2}{1-|a|^2}$). Furthermore, applying the conservation rules we have at time $t_0$  $\mathcal{P}_\spadesuit(t_0)=f_\spadesuit\gamma^{(\spadesuit)}_{t_0}=f_\spadesuit |a|^2$, and  $\mathcal{P}_\heartsuit(t_0)=f_\heartsuit\gamma^{(\heartsuit)}_{t_0}=f_\heartsuit |b|^2$~\footnote{Note, however that these probabilities $\mathcal{P}_{\spadesuit/\heartsuit}(t_0)$ should not  be falsly interpreted as telling that the particle has actually a spin component $\spadesuit/\heartsuit$ at time $t_0$. Instead, this is just a fluid conservation property, and the spin at time $t_0$ could be better defined using the Bohm-Schiller-Tiomno theory~\citep{Bohm1955}.}. \\ 
\indent Usually in the PWI we put $f_\spadesuit=f_\heartsuit=1$ corresponding to the quantum equilibrium regime, i.e., Born's rule,  but this is not necessary. For instance, if we select $f_\spadesuit=\frac{|a|^2+|b|^2}{|a|^2}=|a|^{-2}$ and $f_\heartsuit=0$ we have $\mathcal{P}_\spadesuit(t_f)=1$ and $\mathcal{P}_\heartsuit(t_f)=0$ corresponding to a highly non-equilibrated distribution.  Following the WLLN and Cournot-Borel principle,  this implies that on a long-run Bernoulli sequence we have typically $Q_\spadesuit\simeq 1$ and $Q_\heartsuit\simeq 0$, i.e., strongly conflicting with  Born's rule (i.e., $Q_\spadesuit\simeq |a|^2$, $Q_\heartsuit\simeq |b|^2$).\\
\indent Now, in the PWI we can motivate or even justify Born's rule $f_t=1$~\citep{Durr1992,Valentini1992,Callender2007,Drezet2017} since this justification is ultimately connected to a postulate on the initial conditions of the Bohmian particles (or more generally beables) in a deterministic Universe. For a collapse model like GRW~\citep{GRW} this is postulated as a stochastic law for the actual world.  However, for the MWI this postulate is not possible and can not be motivated by any fundamental principle  since everything in this theory is based on the pure unitary evolution $U_t$ and nothing more!\\
\indent   But actually this is even worse: Remember what we said previously about the absence of typicality in the MWI.   There is no way in this interpretation to define univocally a Cournot-Borel condition  for fixing $Q_\spadesuit/\heartsuit$. Since all branches are needed a case like $Q_\spadesuit\simeq 1$ and $Q_\heartsuit\simeq 0$ (that is strongly conflicting with  Born's rule) must be allowed.  And now, we see that the Schr\"odinger theory or the Bohm-Schiller-Tiomno theory~\citep{Bohm1955} completely justifies the existence of probability distributions associated with a regime of non-equilibrium (e.g., $\mathcal{P}_\spadesuit(t_f)=1$ and $\mathcal{P}_\heartsuit(t_f)=0$). With our previous discussion about estimators and confidence intervals,  it is always possible to find values $f_\spadesuit$ and $f_\heartsuit$ for fitting the empirical values $Q^h_\spadesuit$ and $Q^h_\heartsuit$ in the branch $h$ where we are located; even if that branch is a maverick one!~\footnote{The results obtained with the simple branch counting advocated by Putman and others \citep{Putnam2005,Hemmo2007,Bricmont2016} correspond to $f_\spadesuit=\frac{1}{2|a|^{2}}$, $f_\heartsuit=\frac{1}{2|b|^{2}}$ which yields $\mathcal{P}_\spadesuit=\mathcal{P}_\heartsuit=\frac{1}{2}$ selecting branches with frequencies  $Q_\spadesuit\simeq Q_\heartsuit\simeq \frac{1}{2}$ in the topological Bernoulli tree with $2^N$ branches illustrated in Fig~\ref{figure1}.}\\
\indent  Therefore, all the purported attempts to prove Born's rule within the current many-worlds framework are necessarily rooted in circular or misguided reasonings contradicting the mere spirit of the Everettian ontology. In other words, whereas for a `normal' single-world theory (like the PWI or GRW) the probability function $\mathcal{P}_\alpha$ has a univocal and objective meaning, this can not be so in the MWI where only the frequencies $Q^h_\alpha$ can be defined for the history $h$ considered.  Moreover, there is no rule for the convergence  of $Q^h_\alpha$ and all maverick and magical worlds are allowed in conflict with Cournot's principle.   For example, suppose  one is repeating $N_1=10^{10}$ times a quantum toss experiment and then again $N_2=10^{10}$ more times the same experiment.  We could find worlds where we have for the $N_1$ first trials $h_1=[\spadesuit,\spadesuit,... \spadesuit]$  and subsequently $h_2=[\heartsuit,\heartsuit,... \heartsuit]$ for the $N_2$ next trials.  This looks completely odd and atypical but such worlds exist in the MWI. After the first sequence we could fit the frequency $Q^{h_1}_\spadesuit= 1$ with $f_\spadesuit=|a|^{-2}$, $f_\heartsuit=0$,  and the second sequence  $Q^{h_2}_\heartsuit=1$ could be analyzed with $f_\heartsuit=|b|^{-2}$ and $f_\spadesuit=0$ (of course regrouping the two maverick sequences in a single one we get $Q^{h_1\cup h_2}_\spadesuit= 1/2$ and thus  $f_\spadesuit=\frac{1}{2|a|^{2}}$, $f_\heartsuit=\frac{1}{2|b|^{2}}$). Following our application of the strong-Cournot principle  that would support crazy laws of nature that are never really observed and would conflict  with all known empirical quantum facts such as  Born's rule.\\ 
\indent As a second example of crazy feature that can not be rejected in the MWI imagine an experiment with a beam-splitter, like before, where the observer sees $Q^h_\spadesuit\simeq 1$ even though $|a|=|b|=1/\sqrt{2}$. This corresponds, as we explained, to an atypical `maverick' or `crazy' statistics.  If the number of trials  $N$ is very large the confidence level is so high that deviations are universally negligible in the Borel sense.   Now,   imagine that the observer after having obtained  his/her results do a second experiments with a second beam-splitter after the first one for reuniting the two beams (the exit beams of the first beam-splitter are the input beams of the second one). This is a Mach-Zehnder interferometer and by tuning the phase we can force the beam to always exit in one door corresponding to the state $|+\rangle$:  the other door (i.e., $|-\rangle$ ) being always empty. Now, that means that we have a really strange thing   because with the two experiments  (repeated   $N\gg 1$ times on different qubits)  the observer can conclude that i) with the first experiment all the particles exit in the $\spadesuit$ state: which implies that the beam $\heartsuit$ is empty, ii) whereas in the second experiment the particles go always to the $+$ state which is a signature of interference and therefore   seems to imply that something was in both imput branches $\spadesuit$ and $\heartsuit$ (in the PWI that would be the signature of an empty-wave). Again, this is a consequence of the strong non-equilibrium recorded in the `crazy' observer branch~\footnote{An example of the same kind could be done with the which-path and double-slit experiments.  A first experiment would reveal that all particles are going to slit $1$, whereas a second experiment would reveal interference fringes characteristic of a wave involving the two slits $1$ and $2$.}!        Finally, that means that observers in different branches of the Bernoulli tree can define their ow domain of typicality with their own $f_t$ function.  Therefore, in Everett's relative state theory, i.e., the MWI, probabilities can not have a objective meaning:  probabilities are relative and consequently badly defined (in strong conflict with  Cournot's principle).  And don't forget that Born's rule, as we know it, is very robust empirically since the strong Cournot principle doesn't allow significant deviations for $N$ large. But again, if we accept the usual MWI there is nothing that explains why we don't see maverick worlds: This, we believe, irrevocably constitutes a dead-end for the MWI unless we are going to modify the theory as suggested by several authors.
        
\section{Conclusion}
\indent To conclude this analysis it is important to go back to Everett himself. In his PhD thesis Everett provided a motivation for introducing a typicality reasoning. As he wrote:        
 \begin{quote}
\textit{We wish to make quantitative statements about the relative frequencies of the different possible results of observation-which are recorded in the memory= for a typical observer state; but to accomplish this we must have a method for selecting a typical element from a superposition  of orthogonal states.}~\citep[p.~70]{DeWitt1973}
\end{quote}  As we showed in our analysis of the typicality concept this is the point where he failed. \\
\indent As we showed in the usual probability calculus method the notion of typicality acquires a physical meaning through the strong Cournot principle advocated by Borel. Following this principle to apply the probablity calculus we must define a regime of universally negligible probability or of atypicality. Without this concept it is impossible to transform a potentiality into actuality and the probability calculus would remain unfalsifiable and useless. Many objections raised in the past against the relevance of probablity for science were based on inapropriate applications of such a Cournot principle.\\ 
\indent Moreover, in the MWI one of the subtle point often challenged is that all alternatives occur in decohered or orthogonal branches.  Therefore, the concept of actuality seems to be in jeopardy and the standard Laplace notion of probability applied to alternatives has been criticized. This confusion has certainly inconscioulsy be used by many advocates of the MWI to justify a new and specific concept of probablity used in the theory and to refute criticisms. However, Everett was relatively clear on that point. The introduction of the observer as an automaton with memory of only one of the numerous branches of the Bernoulli tree removes potential ambiguity and connects the standard probability method for alternatives (i.e., as used in single-world theories) to the apparently different situation occuring in the MWI. As he wrote:  
\begin{quote}
\textit{We have, then, a theory which is objectively causal and continuous, while at the same time subjectively probabilistic and discontinuous.}~\citep[p.~69]{Barrett2012}
\end{quote}
In other words the branching tree of life of the observer memory provides the so called `illusion of probability' advocated by Vaidman.  However, as we showed this subjective perspective is not enough to emulate all aspects of the standard probability calculus. The illusion of probability is not enough to give us quantitative predictions for reproducing frequencies.  Indeed, the success of the probability calculus relies on the unambiguous application of the WLLN itself relying on the strong Cournot principle.  This methods shows that we can use unambiguously probabilities as estimators of frequencies for all practical purposes.  The probabilistic convergence of frequencies $Q_\alpha$  to $\mathcal{P}_\alpha$ is postulated to be so efficient that never we can see atypical maverick histories or branches in our real world.  By `real', we mean the empirical evidences provided by standard quantum mechanics imposing Born's rule univocally. However,  this is not possible in the bare MWI and no subjective/Bayesian notion of probability associated with the observer could ever changes that matter of fact. As we showed, with our example of a quantum coin tossing,  the fractal Bernoulli tree  with $2^N$ branches is coherent with infinitely many measures of probability and typicality $\mathcal{P}_\spadesuit$. According to the WLLN, each sectors of typicality selects a frequency $Q_\spadesuit$ in the limit $N$ large and therefore all maverick worlds have their own rationality and justification in the MWI. It would be wrong to say, as many advocates of the MWI nevertheless do, that the situation with the MWI is not worse than in other interpretations of quantum mechanics, or like Everett, that `the situation is fully analogous to that of classical statistical mechanics'~\citep[p.~125]{Barrett2012}. The situation could not actually be more different! In every single world models (or even in the many Bohm-de Broglie worlds ontology) reproduding quantum mechanics the application of Cournot's principle and of the WLLN leads to a clean univocal definition of typicality. Like in classical statistical mechanics this is mandatory in order to apply the probabilityy calculus method.\\
\indent Furthermore, the measure of typicality Everett introduced is far from being univocal as we showed with our analysis of the de Broglie-Bohm theorem $\frac{d}{dt}f_t=0$. As we analyzed the role of the $f_t$ function often neglected provides the most general framework for discussing (continuously evolving) probability distributions in quantum mechanics.  This refutes the claim of a natural and univocal measure of probability imposed by symmetries or anything else in the bare version of the MWI.\\
\indent Finally, to close on a positive point we believe that this analysis provides clues and motivations  for developing other ontologies based on the MWI.  By adding a Born-Vaidman postulate it is already shown that the old MWI can be saved.  However, we believe that this can only be done if we provide a `physical explanation' for the additional principle. By physical explanation we don't only mean  that the principle should reproduce and justify all empirical evidences like Born's rule and locality (that is the road followed currently by Vaidman~\citep{Vaidman2020}), but also that this principle should find it explanatory power in the common and accepted methods of theoretical physics. These accepted explanatory recipes tell that if the physical theory has to be deterministic then the new explanations should either comes from new evolution equations (may be nonlinear~\citep{DeWitt1971,Hanson2003,Buniy2006}), specific boundary or initial conditions (like in the PWI or the many de Broglie-Bohm worlds approach \citep{Tipler2014,Bostrom2014,Sebens2014}), or perhaps by adding some structures (like observers) exploiting the pseudo-randomness associated with initial conditions in the deterministic Universe~\citep{Drezet2021}. One of this strategy could perhaps provide an atypical path for making sense of Everett's program.   
 
\end{document}